\documentclass[twocolumn,aps,floats,letterpaper,floatfix,groupedaddress]{revtex4-2}
\usepackage{float}
\usepackage{afterpage}
\usepackage{amsmath}
\usepackage{amsfonts}
\usepackage{dsfont} 
\usepackage{amssymb}
\usepackage{makeidx}
\usepackage{graphicx}
\usepackage{subfig}
\usepackage{pgfplots}
\usepackage{array}
\usepackage{physics}
\usepackage{color}
\usepackage{mathptmx}
\usepackage{changes,enumitem,todonotes}
\usepackage{soul}
\usepackage[mathscr]{euscript}
\usepackage[colorlinks=true,linkcolor=blue,urlcolor=blue,citecolor=blue]{hyperref}
\usepackage{tikz}
\usepackage[normalem]{ulem}
\usepackage[utf8]{inputenc}
\usepackage{mathtools} 
\usepackage{graphicx}
\usepackage{lipsum} 
\usepackage[colorlinks=true,
  linkcolor=blue,
  citecolor=blue,
  urlcolor=blue]{hyperref}

\begin{document}

\title{From Chiral Topological Dynamics to Chiral Topological Amplification: Real vs Imaginary Parameters in a Hermitian Bosonic Chain}

\author{Kiran Babasaheb Estake\(^{1}\), T. R. Vishnu\(^{1}\), and Dibyendu Roy\(^{1}\)}
\affiliation{\(^{1}\) Raman Research Institute, Bengaluru 560080, India}

\begin{abstract}
We propose a Hermitian quadratic bosonic model (QBH) whose dynamical matrix exhibits distinct topological and dynamical phenomena depending on whether the hopping and pairing amplitudes are real or purely imaginary. In the real-parameter regime, the dynamical matrix is unitarily equivalent to four decoupled copies of the sublattice-symmetric non-Hermitian Su–Schrieffer–Heeger (nSSH2) model, thereby inheriting its topological phases and energy spectrum—including the M{\"o}bius phase, a gapless topological phase with fractional winding number, having no Hermitian counterpart. We show that the dynamics generated by the QBH Hamiltonian naturally reproduce non-Hermitian time evolution, without invoking nonlinear Schr{\"o}dinger dynamics or ad hoc normalization. It is demonstrated by analytically calculating the Loschmidt amplitude and computing the dynamical topological order parameter under periodic boundary conditions, which displays a distinct chiral response in the M{\"o}bius phase. In contrast, when the hopping and pairing terms are taken to be purely imaginary, the dynamical matrix becomes unitarily equivalent to a different version of the sublattice-symmetric non-Hermitian Su–Schrieffer–Heeger (nSSH1) model that supports only two topological phases: trivial and non-trivial, and the M{\"o}bius phase disappears. The latter system exhibits sublattice-dependent chiral amplification under open boundary conditions. We show that this amplification arises from the non-trivial topology of the dynamical matrix, establishing a clear link between topological phase and amplification behavior in the imaginary-parameter regime.
\end{abstract}
\maketitle

\section{Introduction}
Non-Hermitian physics has opened new frontiers in the study of topological phases of matter \cite{Kawabata2019a,Bergholtz2021,Yokomizo2019,Wanjura2020,Borgnia2020,Zhang2020,Okuma2023}. While Hermiticity has long been a cornerstone of quantum theory, ensuring real energy spectra and unitary evolution, realistic systems are often not perfectly isolated, and their effective descriptions can become non-Hermitian due to coupling with the environment or gain and loss mechanisms. In recent years, non-Hermitian topological systems have revealed striking phenomena, including the breakdown or modification of bulk-boundary correspondence (BBC) \cite{Lee2016}, the emergence of the non-Hermitian skin effect (NHSE) \cite{Yao2018, Okuma2020,Zhang2022}, the exceptional points (EPs) \cite{Heiss2012}, and a proliferation of topological phases beyond the Hermitian classification \cite{Gong2018, Kawabata2019, Lieu2018(1)}. Recent studies have shown that non-Hermitian Hamiltonians can emerge naturally in the dynamics of Hermitian quadratic bosonic Hamiltonians (QBHs) that do not conserve particle number, even in the absence of dissipation \cite{Wang2019, Slim2024}. This opens up an intriguing avenue for exploring non-Hermitian physics within strictly Hermitian systems. For instance, the bosonic Kitaev chain—a Hermitian QBH with imaginary hopping and pairing terms was shown to possess a dynamical matrix unitarily equivalent to the Hatano-Nelson model \cite{Hatano1997}, a paradigmatic non-Hermitian system with asymmetric hopping. 

\begin{figure}[h]
  \centering
  \includegraphics[width=\columnwidth]{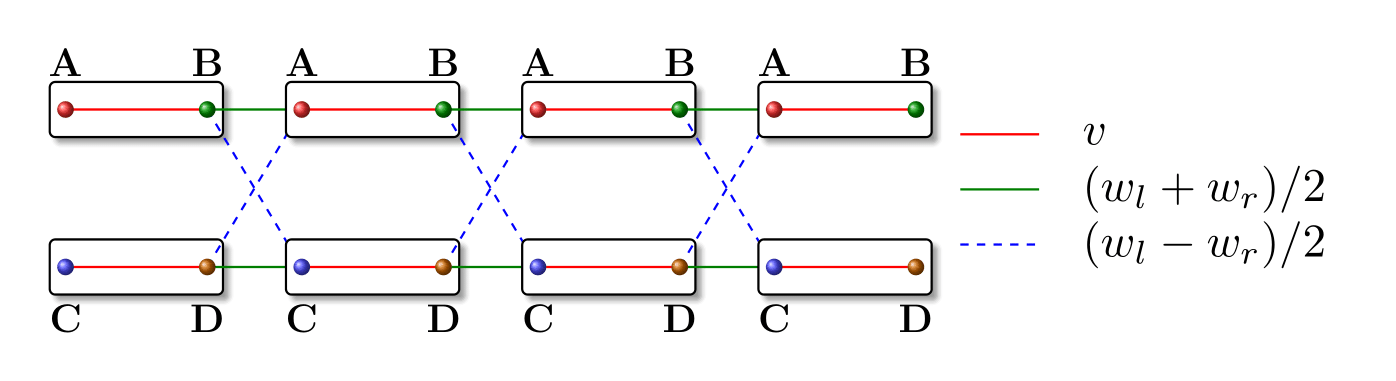}
  \vspace{0.8em}
  \large
  \renewcommand{\arraystretch}{1.3}
  \resizebox{\columnwidth}{!}{%
  \begin{tabular}{|>{\centering\arraybackslash}m{3.5cm}|>{\centering\arraybackslash}m{4.2cm}|>{\centering\arraybackslash}m{4.2cm}|}
  \hline
  \textbf{Feature} & \textbf{Real parameters} & \textbf{Imaginary parameters} \\
  \hline
  $\nu=1/2$ phase & Present & Absent \\
  \hline
  BBC & Absent & Present \\
  \hline
  NHSE & Present & Absent \\
  \hline
  Chiral amplification & Absent & Present \\
  \hline
  Chiral DTOP & Present & Absent \\
  \hline
  \end{tabular}%
  }
  \caption{Top: Cartoon illustration of the QBH model as two copies of Hermitian SSH chains with intra-chain hopping (solid lines) and inter-chain pairing (dotted lines). Bottom: Comparison of various phenomena in real and imaginary parameter regimes of the model.}
  \label{fig:qbh_model_combined}
\end{figure}

In this work, we propose a generalized version of the bosonic Kitaev chain— a Hermitian QBH whose dynamical matrix is unitarily related to the sublattice-symmetric non-Hermitian Su-Schrieffer-Heeger (nSSH) model. The model consists of four sublattices labeled $A$, $B$, $C$, and $D$, with staggered hopping and inter-sublattice bosonic pairing terms that break particle number conservation (see Fig.~\ref{fig:qbh_model_combined}). Remarkably, depending on whether the hopping and pairing amplitudes are real or purely imaginary, the dynamical matrix realizes different versions of the nSSH model, each with distinct topological properties. When the parameters are real, the dynamical matrix is unitarily equivalent to four decoupled copies of the nSSH2 model—a version of the nSSH model \cite{Lee2016, Yao2018, Yin2018, Vyas2021, Nehra2024}, with two EPs, supporting all three known phases under periodic boundary condition (PBC): a gapped trivial phase (winding number $\nu=0$), a gapped topological phase ($\nu=1$), and a gapless M{\"o}bius phase characterized by fractional winding number $\nu=1/2$. 
Under open boundary condition (OBC), the Möbius phase vanishes and the NHSE emerges, breaking the BBC. The M{\"o}bius phase exhibits a distinct chiral response in the dynamical topological order parameter (DTOP). In contrast, when the parameters are purely imaginary, the dynamical matrix is unitarily related to a different version of the nSSH model— nSSH1, which has one EP and in which the M{\"o}bius phase collapses into a gap-closing point separating two topologically distinct gapped phases. In this regime, the BBC is restored and the NHSE disappears. Remarkably, the imaginary-parameter regime also gives rise to sublattice-dependent directional amplification—a phenomenon absent in the real-parameter regime. We analyze the two regimes in detail. In the real-parameter case, we focus on the Möbius phase, which exhibits an anomalous chiral response following a sudden quantum quench \cite{Nehra2024}. While quench dynamics in Hermitian \cite{Heyl2013, Vajna2014, Budich2016, Heyl2018} and non-Hermitian \cite{Zhou2018, Zhou2021, Mondal2022, Mondal2024, Sim2023, Lin2025, Lin2025(1)} fermionic systems have been studied extensively, analogous studies in bosonic models remain limited. Here, we perform a sudden quantum quench of the QBH with real parameters and analytically compute the Loschmidt amplitude. The resulting expression precisely matches that of the nSSH2 model (apart from an overall normalization factor commonly introduced to conserve norm in non-Hermitian evolution \cite{Mondal2022, Zhou2018, Nehra2024}). This match is nontrivial: while the nSSH2 model is fermionic and total particle number is conserved, our QBH is bosonic and includes pairing—breaking total particle number conservation. The unitary equivalence of the dynamical matrix to the nSSH2 model shows that the Loschmidt amplitude in our model inherits its structure entirely from the nSSH2 model, leading to identical expressions for both.

In contrast, the model's dynamical matrix differs significantly in both spectrum and topology for purely imaginary parameters. We show that the matrix is unitarily related to a different version of the sublattice-symmetric non-Hermitian SSH model (nSSH1), which lacks the M{\"o}bius phase but exhibits directional amplification. The amplification is found to be sublattice-dependent and topological in origin \cite{Slim2024}, i.e., it appears only in the non-trivial topological phase of the model. 
Both the magnitude and phase of the hopping and pairing terms can be controlled via the amplitude and phase of external drives in various bosonic platforms, such as optomechanical arrays and superconducting quantum circuits \cite{Slim2024, Busnaina2024}, providing direct experimental access to the real and imaginary parameter regimes.
Finally, our model shares structural similarities with the bosonic Kitaev chain, which has already been realized experimentally \cite{Slim2024, Busnaina2024}.  It suggests that our model, too, is experimentally accessible, opening exciting prospects for probing non-Hermitian topological phases and dynamics in bosonic platforms.

The rest of the paper is organized as follows. In Sec. \ref{real_regime}, we introduce and analyze our proposed model in the real parameter regime. In Sec. \ref{NHSSH}, we review the nSSH2 model and its topological phases. We then present our QBH in Sec. \ref{QBH}, discussing its dynamical matrix, symmetries, and topological phases. In Sec. \ref{QBHresp}, we compute the Loschmidt echo analytically for our QBH and present results for the Pancharatnam geometric phase, return rate, and the DTOP. In Sec. \ref{imag_regime}, we present the analysis of our model in a purely imaginary parameter regime. We discuss the dynamical matrix's spectral properties and topological phases in Sec. \ref{ImDynMat}. In Sec. \ref{amplification}, we demonstrate the sublattice-dependent chiral amplification and establish a connection between this amplification and the topology of the dynamical matrix. Finally, we discuss the potential physical realizations of our models in Sec. \ref{Physical_realization}. We conclude with a summary and outlook in Sec. \ref {sum}. Details of the derivations are provided in six appendices to maintain the flow of the main text. Derivation of the winding number formula is provided in App.~\ref{App.WN}. In App.~\ref{NH_dyn_frm_Herm_QBH}, we discuss how effective non-Hermitian dynamics emerge in QBH systems. The derivation of the Loschmidt amplitude is presented in App.~\ref{der_gk}. Amplification in the imaginary parameter regime is discussed in App.~\ref{App.susc}, while App.~\ref{LE_for_imaginary} discusses the Loschmidt echo and the associated dynamical response in the DTOP for the same regime. Finally, amplification in the real parameter regime is addressed in App.~\ref{amp_for_real}.

\section{The quadratic bosonic model in the real parameter regime}\label{real_regime}

In this section, we introduce our quadratic bosonic model with real-valued hopping and pairing amplitudes and analyze its spectral and topological properties. To understand the structure of the resulting dynamical matrix, we show that it is unitarily equivalent to four copies of nSSH2 chain. We therefore begin by reviewing the nSSH2 model and its topological phases, which will serve as a useful reference throughout our analysis.

\subsection{Non-Hermitian SSH model and its topology}\label{NHSSH}
The Su-Schrieffer-Heeger (SSH) model has been one of the simplest and paradigmatic models of one dimensional (1D) topological insulators, both Hermitian as well as non-Hermitian \cite{SSH}. Here, we discuss a sublattice symmetric version of the non-Hermitian SSH (nSSH2) model studied in \cite{Yao2018, Lieu2018,Vyas2021,Nehra2022}. The real space Hamiltonian of the nSSH2 model is given by

\begin{align}
\hat{H}_{nSSH2} &= \sum_{n =1}^{N} v \hat{c}_{n,A}^{\dagger} \hat{c}_{n,B} + v \hat{c}_{n,B}^{\dagger} \hat{c}_{n,A}\notag\\
 &+ \sum_{n=1}^{N-1} \left(w_r \hat{c}_{n+1,A}^{\dagger} \hat{c}_{n,B} + w_l \hat{c}_{n,B}^{\dagger} \hat{c}_{n+1,A}\right),
\end{align}

where $\hat{c}_{n,\alpha}^{\dagger}$ and $\hat{c}_{n,\alpha}$ are the fermionic creation and annihilation operators of an electron at sublattice $\alpha~(=A,B)$, of unit cell $n$, respectively. Here, $v$ is the intracell hopping amplitude from sublattice $A$ to $B$, and $w_r$ and $w_l$ are respectively, right and left intercell hopping amplitudes among sublattice $A$ and $B$. We obtain the momentum-space Hamiltonian by applying PBC and taking Fourier transform of the creation and annihilation operators. The $k$-space Hamiltonian is given by,

\begin{align}\label{HnSSH(k)}
\hat{H}_{nSSH2}&=\sum_k\hat{\Psi}_k^\dagger H_{nSSH2}(k) \hat{\Psi}_k,\notag\\
H_{nSSH2}(k)&=\begin{pmatrix}0&v+w_re^{-ik}\\v+w_le^{ik}&0\end{pmatrix},
\end{align}

where $\hat{\Psi}_k^\dagger=\begin{pmatrix} \hat{c}^\dagger_{k,A}& \hat{c}^\dagger_{k,B}\end{pmatrix}$.  We parameterize $v=J(1-\delta)$ and $w_r=w_le^{-\theta}=J(1+\delta)$. So, the parameter $\theta$ controls non-Hermiticity, i.e., we get the Hermitian SSH model if $\theta=0$. The Hamiltonian matrix $H_{nSSH2}(k)$ can be expressed in terms of the Pauli matrices $\sigma_i,$ $i=x,y,z$ as $H_{nSSH2}(k)=\vec{d}(k)\cdot \vec{\sigma}$, where the Bloch vector $\vec{d}(k)$ has real and imaginary components since $H_{nSSH2}(k)$ is non-Hermitian. Let $\vec{d}(k)=\vec{d}^r(k)+i\vec{d}^i(k)$, where the real and imaginary parts of $\vec{d}(k)$ are denoted by $\vec{d}^r(k)$ and $\vec{d}^i(k)$, respectively. These are given by

\begin{align}\label{d_vec}
\vec{d}^r(k)&=\Bigl(v+\frac{(w_l+w_r)\cos k}{2}\Bigr)\hat{x}+\frac{(w_l+w_r)\sin k}{2}\hat{y},
\notag\\
\vec{d}^i(k)&=\frac{(w_l-w_r)\sin k}{2}\hat{x}-\frac{(w_l-w_r)\cos k}{2}\hat{y}.
\end{align}

\begin{figure}[h]
  \centering
 {\includegraphics[width=0.45\textwidth]{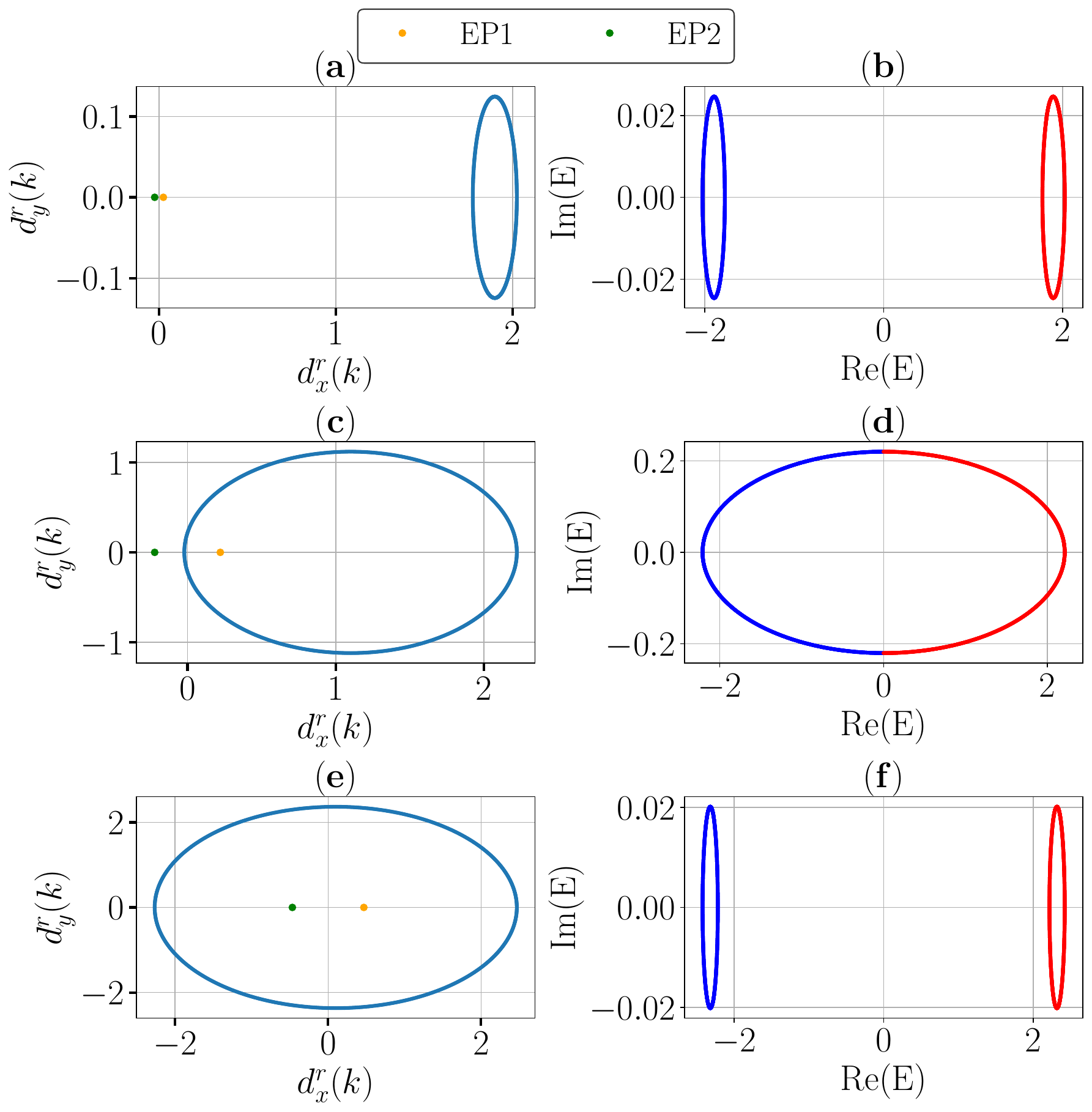}}
  \caption{Schematic diagram of three different winding numbers and the corresponding parametric energy plots of $H_{nSSH2}(k)$. (a) $\delta=-0.9$, no exceptional point is enclosed by the closed curve, $\nu_1=\nu_2=\nu=0$. (b) The complex energy bands form two isolated closed loops with real energy gap. (c) $\delta=-0.1$, EP1 is enclosed while EP2 is not, $\nu_1=1$, $\nu_2=0$ which makes $\nu=1/2$. (d) The two energy loops come close together and merge to form a single bigger loop and there is no real energy gap. (e) $\delta=0.9$, both the exceptional points are enclosed, $\nu_1=\nu_2=1$ which makes $\nu=1$. (f) The two loops again separate from each other and there is a real gap in between them. We take $J=1$ and $\theta=0.4$ in (a)-(f).}
  \label{WN_Para_plots_H_nSSH}
\end{figure}

Both vectors are restricted on the $x$-$y$ plane as a consequence of sublattice symmetry, which is described as $\sigma_zH_{nSSH2}(k)\sigma_z^{-1}=-H_{nSSH2}(k)$. Another consequence of sublattice symmetry is that the energies $E(k)$ of the Hamiltonian come in pairs, $E_+(k)$ and $E_-(k)$, given by $E_{\pm}(k)=\pm\sqrt{v^2+w_rw_l+v(w_le^{ik}+w_re^{-ik})}$. These energies define two bands in the complex energy plane. The points where the bands touch each other are called the EPs. At these points, $E_{\pm}=0$ which happens at $k=\pi$ and when $v^2-v(w_l+w_r)+w_lw_r=0$. This is a quadratic equation in $v$ and has two roots, $v=w_l$ and $v=w_r$. So there are two EPs for this model. It can be shown that for two band sublattice symmetric 1D models, like the nSSH2 model, the winding number is defined by just $\vec{d}^r(k)$ alone (see App. \ref{App.WN}). Let us denote the exceptional points in the ${d_x}^{r}$-${d_y}^{r}$ plane as EP1 and EP2. These are given by EP1: $\vec{d}^r(k)=\frac{1}{2}(w_l-w_r)\hat{x}$ and EP2: $\vec{d}^r(k)=-\frac{1}{2}(w_l-w_r)\hat{x}$. It can be understood by taking $k=\pi$ in Eq. \ref{d_vec} and by substituting $v=w_r$ and $v=w_l$ separately.  The same EPs occur at $k=0$ too when $v=-w_r$, and $v=-w_l$. The winding number turns out to be $\nu=\frac{1}{2}(\nu_1+\nu_2)$ (see App. \ref{App.WN}) \citep{Yin2018}, where $\nu_1$  ($\nu_2$) is the number of times  $\vec{d}^r(k)$ encircles the EP1 (EP2) in ${d_x}^{r}$-${d_y}^{r}$ plane (see Fig. \ref{WN_Para_plots_H_nSSH}). The nSSH2 model has three topological phases. When $\delta<\frac{1-e^\theta}{1+e^\theta}$, it can be shown that $\nu_1=\nu_2=0$, so $\nu=0$. In the region where $\frac{1-e^\theta}{1+e^\theta}<\delta<0$, we get $\nu_1=1$ and $\nu_2=0$, so $\nu=1/2$. And when $\delta>0$, we have $\nu_1=\nu_2=1$ giving us $\nu=1$. These are demonstrated in Figs. \ref{WN_Para_plots_H_nSSH}(a), (c) and (e). The complex energy plots corresponding to these winding numbers are also shown in Figs. \ref{WN_Para_plots_H_nSSH}(b), (d) and (f), respectively. The two bands are shown in red and blue, as $k$ varies from $-\pi$ to $\pi$ in the Brillouin zone (BZ). The gapless phase with fractional winding number in the nSSH2 model does not have a Hermitian analog. This phase is observed to have features similar to the M{\"o}bius strip \cite{Vyas2021}. We will refer to this phase as the Möbius phase. It exhibits an anomalous chiral response in the DTOP \cite{Nehra2024}. 
Notably, the M{\"o}bius phase is absent under OBC, a feature that can be interpreted as a manifestation of the breakdown of the BBC. 

In the Hermitian limit ($w_r=w_l=w$), the imaginary part of the Bloch vector, $\vec{d}_i(k)$ vanishes and the exceptional points EP1 and EP2 merge into one point, the origin of the ${d_x}^{r}$-${d_y}^{r}$ plane. So, the winding number of the Hermitian SSH model is the number of times the Bloch vector winds around the origin. It turns out that, for Hermitian SSH model, $\nu=0$, when $v>w$ and $\nu=1$ when $v<w$ \citep{Asboth}.

\subsection{Quadratic Bosonic Hamiltonian}\label{QBH}

The QBHs can give rise to non-Hermitian dynamical matrices, even when the Hamiltonian itself is Hermitian. This non-Hermiticity emerges when the system lacks U(1) symmetry and the total particle number is not conserved~\cite{Flynn2020}. We provide a general discussion of this phenomenon in App.~\ref{NH_dyn_frm_Herm_QBH}. Here, we introduce our model, discuss its dynamical matrix, energy spectrum, symmetries, and topological phases in the regime of real hopping and pairing amplitudes. 

Given a non-Hermitian matrix, there exists a mapping from the non-Hermitian matrix to a Hermitian QBH whose dynamical matrix in a certain basis is unitarily related to the given non-Hermitian matrix \cite{Wang2019}. So, it is possible to engineer a Hermitian QBH whose dynamical matrix is related to the given non-Hermitian matrix. Let $\mathcal{H}_N$ be the non-Hermitian Hamiltonian. Then the mapping to the Hermitian QBH is given by \cite{Wang2019}
\begin{align}
\hat{H} &= \frac{1}{2} \sum_{j,j' =1}^{N} \Big[ ({\cal H}_N + {\cal H}_N^{\dagger})_{j,j'} (\hat{a}_j^{\dagger} \hat{a}_{j'} - \hat{b}_j \hat{b}_{j'}^{\dagger}) \notag\\
&+ ({\cal H}_N - {\cal H}_N^{\dagger})_{j,j'} (\hat{a}_j^{\dagger} \hat{b}_{j'}^{\dagger} - \hat{a}_{j'} \hat{b}_j)  \Big],
\end{align}
where $(\mathcal{H}_N)_{j,j'}$ denotes the $(j,j')$-th element of the Hamiltonian matrix $\mathcal{H}_N$.  $\hat{a}_j^{\dagger}, \hat{a}_j$ and $\hat{b}_j^{\dagger}, \hat{b}_j$ are the bosonic creation and annihilation operators of two bosonic modes, say $\alpha$ and $\beta$ at site $j$. This mapping doubles the degrees of freedom in the system, effectively doubling both the system size and the number of sublattice sites per unit cell. We use this mapping to construct a Hermitian QBH whose dynamical matrix is related to the nSSH2 model.  While the nSSH2 model features two sublattices $A$ and $B$, our mapped Hermitian QBH is a four-sublattice model. The Hamiltonian for this Hermitian QBH is given by:

\begin{align}\label{HQB_realspaceham}
\hat{H}_{QB} &=  \sum_{i =1}^{N} v \hat{A}^{\dagger}_i \hat{B}_i +\sum_{i =1}^{N-1}\frac{w_r+w_l}{2}\hat{A}^{\dagger}_{i+1} \hat{B}_i\notag\\
&- \sum_{i =1}^{N} v \hat{C}^{\dagger}_i \hat{D}_i -\sum_{i =1}^{N-1}\frac{w_r+w_l}{2}\hat{C}^{\dagger}_{i+1} \hat{D}_i\notag\\
&+ \sum_{i = 1}^{N-1} \frac{(w_l - w_r)}{2} (\hat{B}_{i}^{\dagger} \hat{C}^{\dagger}_{i+1} - \hat{A}_{i+1} \hat{D}_i) + H. c.,
\end{align}


where, $v$, $w_r$ and $w_l$ are real positive parameters, and $N$ is the number of unit cells, each containing four sublattice sites. The operators $\hat{A}_i$, $\hat{B}_i$, $\hat{C}_i$, $\hat{D}_i$, and $\hat{A}^\dagger_i$, $\hat{B}^\dagger_i$, $\hat{C}^\dagger_i$, $\hat{D}^\dagger_i$, denote bosonic annihilation and creation operators corresponding to sublattices $A$, $B$, $C$, and $D$, respectively. This model can be interpreted as two Hermitian SSH chains—one with sublattices $A$ and $B$, and another with sublattices $C$ and $D$, which are coupled through pairing terms, Fig. \ref{fig:qbh_model_combined}. Importantly, when $w_r=w_l$ (i.e., $\theta=0$), the pairing terms vanish, reducing the system to two independent Hermitian SSH chains. Thus the parameter $\theta$, which governs the non-Hermiticity in the nSSH2 model, now controls the strength of the pairing terms in $\hat{H}_{QB}$. This establishes a direct physical mapping between hopping asymmetry in the non-Hermitian model and the bosonic pairing terms in the Hermitian model.


Applying PBC, we take Fourier transform of the creation and annihilation operators as $\hat{\alpha}_j = (1/\sqrt{N}) \sum_{k =1}^{N} e^{i jk} \hat{\alpha}_k$, where $\hat{\alpha}_j$ can be any of the sublattice operators $\hat{A}_j$, $\hat{B}_j$, $\hat{C}_j$, or $\hat{D}_j$. The $k$-space Hamiltonian takes the form
\begin{align}\label{HQB_k}
\hat{H}_{QB} =& \sum_{k} \Big[ f_1(k) \hat{A}_k^{\dagger} \hat{B}_k-f_1(k)\hat{C}_{k}^{\dagger} \hat{D}_{k} \cr
	& + f^*_2(k) \hat{B}_k^{\dagger} \hat{C}^{\dagger}_{-k} -f^*_2(k) \hat{A}_k \hat{D}_{-k}) \Big] + H. c.,
\end{align}
where the $k$-dependent coefficients are defined as 
\begin{align}\label{f1f2}
f_1(k)&=v+\frac{(w_l+w_r)}{2}e^{-ik},\notag\\
f_2(k)&=\frac{(w_l-w_r)}{2}e^{-ik},
\end{align}
where $v$, $w_r$ and $w_l$ are again parameterized as in nSSH2 model. This Hamiltonian can be recast in the matrix form using the Nambu spinor $\hat{\phi}_k= (\hat{A}_{k}, \hat{B}_{k}, \hat{C}_{k}, \hat{D}_{k},\hat{A}_{-k}^{\dagger}, \hat{B}_{-k}^{\dagger}, \hat{C}_{-k}^{\dagger},  \hat{D}_{-k}^{\dagger})^T$ as
\begin{align}\label{HQB_k2}
\hat{H}_{QB} = \frac{1}{2} \sum_{k} \hat{\phi}_k^{\dagger} H_{QB}(k) \hat{\phi}_k,
\end{align}
where
\begin{align}
 H_{QB}(k)&=\begin{pmatrix}P(k)&Q(k)\\Q(k)&P(k)\end{pmatrix},
\end{align}
where 
\begin{align}
P(k)&=\begin{pmatrix}0&f_1(k)&0&0\\f_1^*(k)&0&0&0\\0&0&0&-f_1(k)\\0&0&-f_1^*(k)&0\end{pmatrix},\notag\\
 Q(k)&=\begin{pmatrix}0&0&0&-f_2(k)\\0&0&f_2^*(k)&0\\0&f_2(k)&0&0\\-f_2^*(k)&0&0&0\end{pmatrix}.
\end{align}
The bosonic commutation relations between the sublattice operators $\hat{A}, \hat{B}, \hat{C}$ and $\hat{D}$ can be compactly expressed in terms of $\hat{\phi}_k$ as 
\begin{align}\label{phi_k_comm}
[\hat{\phi}_k,\hat{\phi}^\dagger_l]=\delta_{kl}\tau_3, \quad \tau_i=\sigma_i \otimes \mathbb{I}_{4}.
\end{align}

\subsubsection{The dynamical matrix}
Using the Heisenberg equation for the creation and annihilation operators, we get the equation for $\hat{\phi}(k)$ as

\begin{align}
i\frac{d}{dt}\hat{\phi}_k(t)=G_{QB}(k)\hat{\phi}_k(t), \quad G_{QB}(k)=\tau_3 H_{QB}(k),
\end{align}
 $G_{QB}(k)$ is the $k$-space dynamical matrix.

\begin{figure}[h]
  \centering
  {\includegraphics[width=0.49\textwidth]{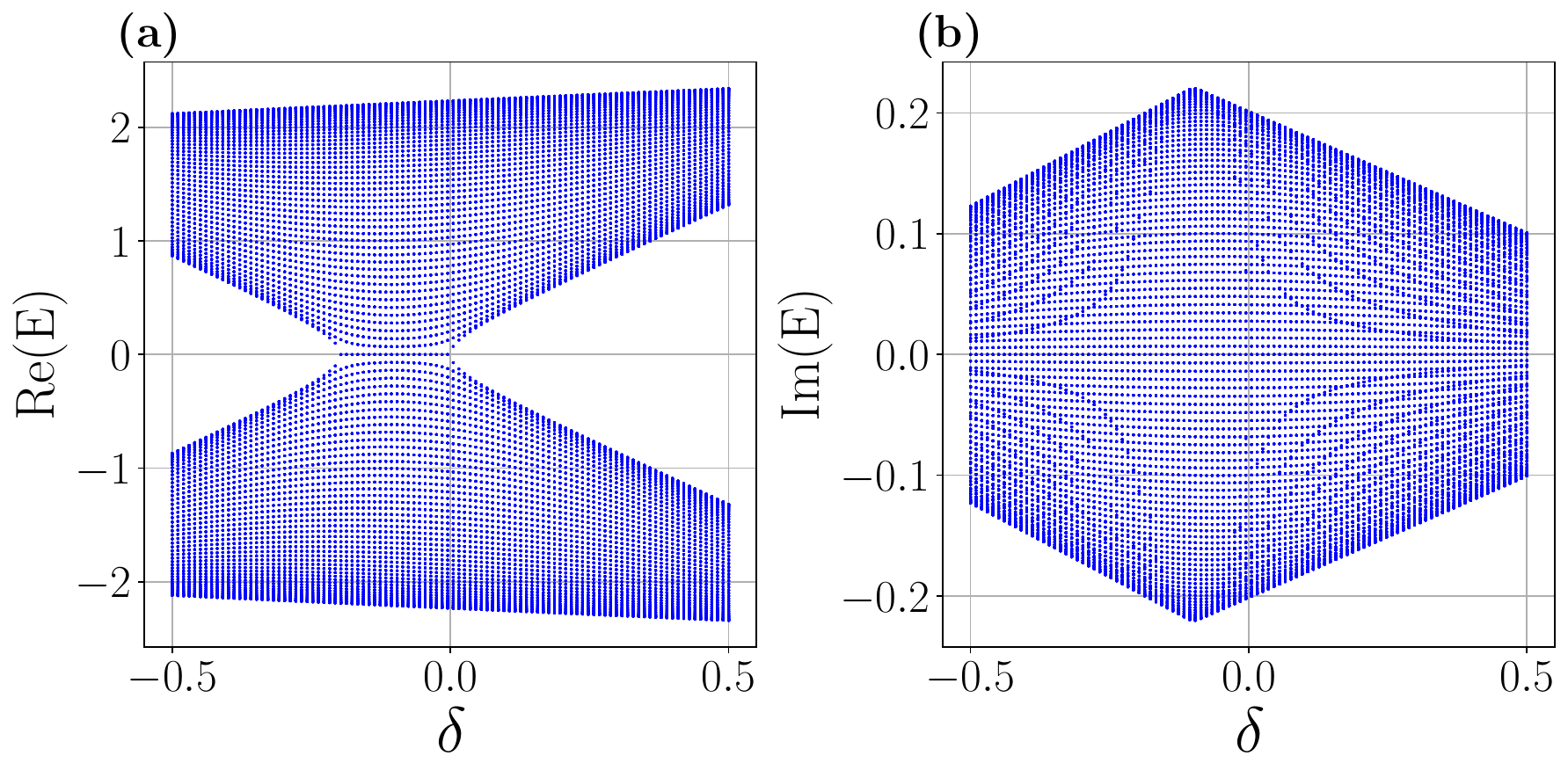}}
  \caption{Real part (a) and imaginary part (b) of the spectrum of $G_{QB}(k)$ as function of $\delta$. We fix $J=1$ and $\theta=0.4$ in both the plots.}
  \label{PBC_spec_with_w}
\end{figure}
We fix $J=1$ and $\theta = 0.4$. Fig. \ref{PBC_spec_with_w} shows the real and imaginary parts of the spectrum of $G_{QB}(k)$ as functions of $\delta$. We observe that in the range $\delta \in (\frac{1-e^\theta}{1+e^\theta},0)$, the real part of the spectrum becomes gapless and certain modes acquire purely imaginary eigenvalues. This region corresponds to the Möbius phase discussed earlier in the context of the nSSH2 model. The presence of purely imaginary energy modes is a hallmark of the Möbius phase. The spectra match exactly with those of the $H_{nSSH2}(k)$ model, up to additional degeneracies. This correspondence and the reason behind degeneracies will be explained in detail through the symmetries of $G_{QB}(k)$ in the next subsection.

\begin{figure}[h]
  \centering
 {\includegraphics[width=0.45\textwidth]{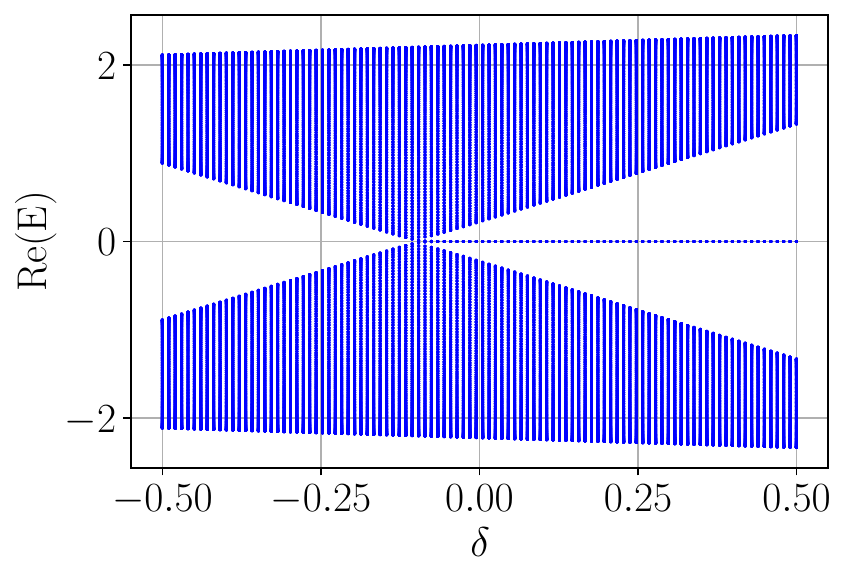}}
  \caption{Spectrum of the dynamical matrix $G_{QB}$ for OBC. It is purely real and there are two gapped phases: trivial phase and non-trivial phase, with zero modes present in the non-trivial phase. The parameters are $J=1$ and $\theta=0.4$.}
  \label{OBC_spec_with_w}
\end{figure}

Fig. \ref{OBC_spec_with_w} displays the spectrum of $G_{QB}$ under OBC. The most striking feature is that the spectrum is entirely real even though $G_{QB}$ is a non-Hermitian matrix. This results from the existence of a similarity transformation that maps the non-Hermitian dynamical matrix under OBC to a Hermitian one—a transformation that is not possible under PBC  \cite{Yao2018}. A similar phenomenon occurs in the $H_{nSSH2}$ model under OBC. Additionally, the Möbius phase appears to vanish in the OBC spectrum. All the eigenvectors get localized at the boundary. This is a manifestation of the NHSE: the pronounced sensitivity of the eigenvalue spectrum and eigenstates to the choice of boundary conditions. It is now well established that topological phases in non-Hermitian systems can differ significantly between PBC and OBC \cite{Lee2016, Okuma2023}, and that the conventional BBC often requires modification in such contexts \cite{Yao2018, Kunst2018}.\\

\textcite{McDonald2018} studied a quadratic bosonic chain, a bosonic analogue of the $p$-wave topological superconducting chain proposed by Kitaev \cite{Kitaev2001}. The bosonic Kitaev chain features nearest-neighbour hopping and pairing terms, and its dynamical matrix is generally non-Hermitian. Depending on the relative strengths of hopping and pairing, the system can exhibit stable behaviour (real energies) or unstable behaviour (complex energies). In contrast to the bosonic Kitaev chain, our model is unstable whenever $w_r\neq w_l$ because it contains only inter-chain pairing and no hopping (see Fig. \ref{fig:qbh_model_combined}). The stability of our model can be restored by introducing inter-chain hopping terms alongside the pairing terms. Similar stability–instability transitions have been studied in other bosonic settings, for example, in open coupled scalar field theories \cite{Vishnu2024}.
 
\subsubsection{Symmetries of the dynamical matrix}\label{QBHsym}
By construction, the dynamical matrix $G_{QB}(k)$ has the particle-hole symmetry (PHS1) and the pseudo-Hermiticity which are given as $\tau_1G_{QB}^*(-k)\tau_1=-G_{QB}(k)$ and $\tau_3G_{QB}^\dagger(k)\tau_3=G_{QB}(k)$ respectively. PHS1 and pseudo-Hermiticity together imply that the eigenvalues of $G_{QB}(k)$ come in pairs $\{E(k),E^*(k),-E(k),-E^*(k)\}$. 
In addition to PHS1, $G_{QB}(k)$ has another particle-hole symmetry PHS2, given by $\tilde{\tau}_1 G_{QB}^*(-k)\tilde{\tau}_1=-G_{QB}(k)$, where $\tilde{\tau}_1=\mathbb{I}_{2}\otimes \sigma_1 \otimes \mathbb{I}_{2}$. The two particle-hole symmetries PHS1 and PHS2 can be combined to construct a unitary symmetry $U=\tau_1\tilde{\tau}_1$ such that $UG_{QB}(k)U^{-1}=G_{QB}(k)$. This unitary symmetry $U$ is important in understanding why the spectra of $G_{QB}(k)$ and $H_{nSSH2}$ match so well. Since $G_{QB}(k)$ has a unitary symmetry $U$, we can transform $G_{QB}(k)$ in block diagonal form. Let $Q$ be the matrix of transformation made out of eigenvectors of $U$. $Q$ is a unitary matrix given by

\begin{align}
Q=\frac{1}{\sqrt{2}}\left(
\begin{array}{cccc}
 I & 0 & 0 & I  \\
 0 & I & I & 0  \\
 0 & -I & I & 0  \\
 I & 0 & 0 & -I  \\
\end{array}
\right),
\end{align}

where $I$ is an identity matrix of size $2$. Let $Q^\dagger G_{QB}(k) Q=G_b(k)$ be the transformed dynamical matrix. $G_b(k)$ turns out to be

\begin{align}\label{Block_GB}
G_b(k)=\left(
\begin{array}{cccc}
 H_{nSSH2}(k) & 0 & 0 & 0  \\
 0 & -H_{nSSH2}^\dagger(k) & 0 & 0  \\
 0 & 0  & -H_{nSSH2}(k) & 0  \\
 0 & 0 & 0 & H_{nSSH2}^\dagger(k)  \\
\end{array}
\right).
\end{align}
This form reveals that $G_{QB}(k)$ is unitarily equivalent to four decoupled nSSH2 chains, represented by $H_{nSSH2}(k)$, $H_{nSSH2}^\dagger(k)$, $-H_{nSSH2}(k)$, and $-H_{nSSH2}^\dagger(k)$. If $E(k)$ and $-E(k)$ are eigenvalues of $H_{nSSH2}(k)$, then the eigenvalues of $H_{nSSH2}^\dagger(k)$ are $E^*(k)$ and $-E^*(k)$. The additional blocks with negative signs simply add degeneracies without altering the spectral shape. This explains both the spectral matching between $G_{QB}(k)$ and $H_{nSSH2}(k)$, and the observed degeneracies in $G_{QB}(k)$.



\subsubsection{Topological phases of the dynamical matrix}\label{QBHtop}
To study the topological aspects of the dynamical matrix, we consider its block diagonal form in Eq. \ref{Block_GB}. Since the blocks correspond to the nSSH2 model, the topological features of the dynamical matrix are expected to mirror those of the nSSH2 model. Let us focus on the block $H_{nSSH2}(k)$ given in Eq. \ref{HnSSH(k)}. The winding number is determined by the real part of $\vec{d}(k)$ \cite{Yin2018}. We derive the winding number formula for this matrix in App. \ref{App.WN}. It is given by
\begin{align}
\nu=\frac{\nu_1+\nu_2}{2},
\end{align}
where
$\nu_{1,2}$ are the winding numbers associated with $\vec{d}^r(k)$ and the two exceptional points EP1 and EP2 given in Sec. \ref{NHSSH}. Specifically, $\nu_1$ and $\nu_2$ count how many times the real part of $\vec{d}(k)$ winds around the exceptional points EP1 and EP2, respectively, as described in Sec. \ref{NHSSH} \cite{Yin2018}. For the parametrization that we are considering, there are three different phases: $\nu=0$ for $\delta<\frac{1-e^\theta}{1+e^\theta}$, $\nu=0.5$ when $\frac{1-e^\theta}{1+e^\theta}<\delta<0$, and $\nu=1$ when $\delta>0$. The different possible winding numbers are illustrated in first column of Fig. \ref{WN_Para_plots_H_nSSH}.

\subsection{Quench dynamics and dynamical topological order parameter}\label{QBHresp}
In quench studies of the nSSH2 model, the M{\"o}bius phase exhibits a distinct chiral response in terms of the Pancharatnam geometric phase and the DTOP \cite{Nehra2024}. This behavior is qualitatively different from that of the trivial and non-trivial phases. When the Hamiltonian is quenched from the Hermitian trivial phase to the non-Hermitian trivial phase, the return rate remains a smooth function of time, and no DTOP is observed. In contrast, a quench from the Hermitian trivial phase to the non-Hermitian non-trivial phase leads to non-analyticities in the return rate and the emergence of a non-zero DTOP in both the positive and negative halves of the BZ. Remarkably, when the quench is across the Hermitian trivial and the M{\"o}bius phase of the non-Hermitian Hamiltonian, the return rate still exhibits non-analytic behavior, but the DTOP appears only in the positive half of the BZ. This chiral response serves as a characteristic signature of the M{\"o}bius phase \cite{Nehra2024}. Here, we extend this line of investigation to the QBH. To our knowledge, this is the first time that the Loschmidt amplitude, return rate, Pancharatnam geometric phase, and DTOP have been studied in the context of the QBH. This analysis is particularly important because, unlike in the non-Hermitian SSH model, where the Schr{\"o}dinger equation is applied despite the non-Hermitian nature of the Hamiltonian, often requiring ad hoc normalization of the time-evolved state, the QBH is Hermitian. This makes the analysis more controlled and conceptually cleaner, while still capturing the non-Hermitian features through the structure of the non-Hermitian dynamical matrix.

In a previous study \cite{Nehra2024}, quench dynamics were investigated with the initial Hamiltonian restricted to be Hermitian. In the present work, we build on the understanding that non-Hermiticity in the nSSH2 model corresponds to pairing terms in the QBH model. We generalize the quench protocol to include scenarios where the initial and final Hamiltonians contain such pairing terms. A priori, it is not apparent how pairing terms in the initial Hamiltonian would affect the dynamics. We show that the expression for the Loschmidt amplitude retains its form, irrespective of the pairing structure in the initial Hamiltonian. This raises a natural question: why do the quench dynamics resulting from the nSSH2 model, which conserves the total number of fermions, match those of the Hermitian QBH, which does not conserve the total number of bosons? We find that the answer has two parts. First, particle number non-conservation is effectively traded for the pseudo-bosonic modes in the QBH, as discussed in this subsection. Second, since the dynamical matrix $G_{QB}$ is unitarily related to four copies of the nSSH2 model (see Eq. \ref{Block_GB}), the resulting calculations conspire to yield an exact match with the nSSH2 model result (up to a normalization factor). Our analytical calculations with numerical simulations reveal dynamical features that reproduce and extend those observed in the nSSH2 model, demonstrating the robustness of the non-Hermitian signatures within a fully Hermitian framework. We hope this work helps clarify the dynamical role of non-Hermiticity in bosonic systems.

\begin{figure}[h]
  \centering
 {\includegraphics[width=0.49\textwidth]{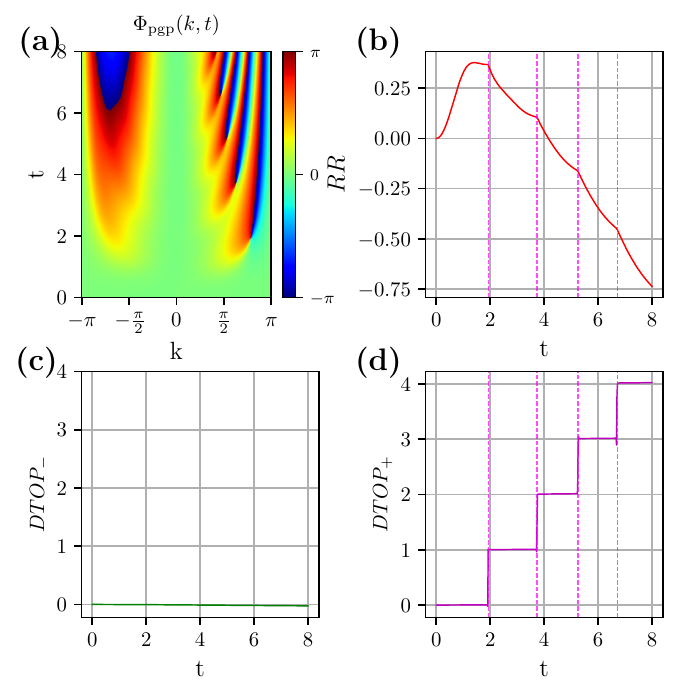}}
  \caption{Dynamical response for the real parameter regime. The Pancharatnam geometric phase, $\phi_{\rm pgp}(k,t)$ in (a), the return rate ($RR(t)$) in (b), the $DTOP_-$ and $DTOP_+$ in (c) and (d), respectively, for the quench from an initial Hamiltonian with parameters $J^i=1$, $\delta^i=-0.9$, $\theta^i=0$ to a final Hamiltonian with parameters $J^f=1$, $\delta^f=-0.1$, $\theta^f=0.4$.}
  \label{t_m}
\end{figure}

The central object in the quench study is the Loschmidt amplitude $G(t)$, defined as the overlap between an eigenstate of the initial Hamiltonian and the time-evolved state under the final (quenched) Hamiltonian. From this quantity, we derive other quantities such as the return rate, the Pancharatnam geometric phase, and the DTOP. To have an analytical handle on the problem, the first essential step is to express the Hamiltonian in Eq. \ref{HQB_k2} in its normal form. This is done as following:
\begin{align}\label{HQB_in_G}
\hat{H}_{QB} = \frac{1}{2} \sum_{k} \hat{\phi}_k^{\dagger} \tau_3 G_{QB}(k) \hat{\phi}_k,
\end{align}
where we have used $\tau_3^2=1$ and $\tau_3H_{QB}(k)=G_{QB}(k)$ in Eq. \ref{HQB_k2}. Since, $G_{QB}(k)$ is a non-Hermitian matrix, its spectral decomposition is carried out using its left and right eigenvectors, which form a biorthogonal basis. Let $|\psi)$ and $(\chi|$ be the notations for the right and left eigenvectors of $G_{QB}(k)$ respectively (we use the rounded ket and bra notations to emphasize that these are not the states in the Hilbert space of Hamiltonian $\hat{H}_{QB}$ but just the left and right eigenvectors of the matrix $G_{QB}(k)$). $G_{QB}(k)$ can be written as
\begin{align}
G_{QB}(k)&=E(k)|\psi^1_+)(\chi^1_+|-E(k)|\psi^1_-)(\chi^1_-|\notag\\
&+E^*(k)|\psi^{1*}_+)(\chi^{1*}_+|-E^*(k)|\psi^{1*}_-)(\chi^{1*}_-|\notag\\
&-E(k)|\psi^2_-)(\chi^2_-|+E(k)|\psi^2_+)(\chi^2_+|\notag\\
&-E^*(k)|\psi^{2*}_-)(\chi^{2*}_-|+E^*(k)|\psi^{2*}_+)(\chi^{2*}_+|,
\end{align}
where $E(k)=\sqrt{v^2+w_rw_l+v(w_le^{ik}+w_re^{-ik})}$. The vectors $|\psi^1_+)$, $|\psi^2_+)$ together with their corresponding left eigenvectors $(\chi^1_+|$, $(\chi^2_+|$, form a set of degenerate right and left eigenvectors of $G_{QB}(k)$ with eigenvalue $E(k)$. Similarly, $|\psi^1_-)$, $|\psi^2_-)$ and $(\chi^1_-|$, $(\chi^2_-|$ correspond to the eigenvalue $-E(k)$. In addition, $|\psi^{1*}_+)$, $|\psi^{2*}_+)$ and $(\chi^{1*}_+|$ and $(\chi^{2*}_+|$ are the degenerate right and left eigenvectors associated with the complex conjugate eigenvalue $E^*(k)$, and likewise for the eigenvalue $-E^*(k)$. Using this decomposition of $G_{QB}(k)$ in Eq. \ref{HQB_in_G} and after doing some simplifications, we get
\begin{align}\label{normal_Hf}
\hat{H}_{QB} &= \sum_{k}E(k){\hat{\eta}_{k+}^{1\dagger}} \hat{\bar{\eta}}^1_{k+}+E(k){\hat{\eta}_{k+}^{2\dagger}} \hat{\bar{\eta}}^{2}_{k+}\notag\\
&-E(k){\hat{\eta}_{k-}^{1\dagger}} \hat{\bar{\eta}}^1_{k-}-E(k){\hat{\eta}_{k-}^{2\dagger}} \hat{\bar{\eta}}^{2}_{k-},
\end{align}
where
\begin{align}
\hat{\eta}_{k\pm}^{1\dagger}&=\hat{\phi}_k^{\dagger} \tau_3|\psi^1_\pm),\notag\\
\hat{\bar{\eta}}^1_{k\pm}&=(\chi^1_\pm|\hat{\phi}_k,\notag\\
\hat{\eta}_{k\pm}^{2\dagger}&=\hat{\phi}_k^{\dagger} \tau_3|\psi^2_\pm),\notag\\
\hat{\bar{\eta}}^2_{k\pm}&=(\chi^2_\pm|\hat{\phi}_k.
\end{align}
The left and right eigenvectors $|\psi^{1,2}_\pm)$ and $(\chi^{1,2}_\pm|$ form a biorthogonal basis and satisfy the following bi-orthonormality relations:
\begin{align}
(\chi^\alpha_\gamma|\psi^\beta_\lambda)=\delta_{\alpha\beta}\delta_{\gamma\lambda},
\end{align}
where $\alpha,\beta \in \{1,2\}$ and $\gamma,\delta \in \{+,-\}$. These conditions, together with the original commutation relations in Eq. \ref{phi_k_comm} lead to the following commutation relations for the normal mode operators $\hat{\eta}_{k\pm}^{1,2\dagger}$ and $\hat{\bar{\eta}}^{1,2}_{k\pm}$:
\begin{align}
[\hat{\bar{\eta}}^{\alpha}_{k\gamma},\hat{\eta}_{l\lambda}^{\beta\dagger}]=\delta_{\alpha \beta}\delta_{\gamma \lambda}\delta_{kl}.
\end{align}
Although these operators obey bosonic commutation relations, they are not Hermitian conjugates or adjoints of each other. Therefore, the pairs $(\hat{\bar{\eta}}^{1,2}_{k\pm},\hat{\eta}_{k\pm}^{1,2\dagger})$  are referred to as pseudo-bosonic normal modes \cite{Flynn2020}. The reason behind this can be traced back to the non-Hermiticity of the dynamical matrix $G_{QB}(k)$, which necessitates the use of biorthogonal left and right eigenvectors. Consequently, physical calculations must carefully account for both left and right states throughout. 
Let $\ket{0}$ be the vacuum of the pseudo-bosonic modes, i.e., 
\begin{align}
\hat{\bar{\eta}}^{1,2}_{k\pm}\ket{0}=0,
\end{align}
and $\bra{\bar{0}}$ be the corresponding left state, such that,
\begin{align}
\bra{\bar{0}}\hat{\eta}_{k\pm}^{1,2\dagger}=0,
\end{align} 
and $\bra{\bar{0}}\ket{0}=1.$
The Loschmidt amplitude is calculated by evolving an eigenstate of the initial Hamiltonian under the dynamics of the final (quenched) Hamiltonian and taking the overlap with the initial state. We define the initial state as
\begin{align}
\ket{\mathcal{G}}=\prod_k \hat{\eta}_{k-}^{1\dagger} \ket{0},
\end{align}
and, its biorthogonal counterpart as
\begin{align}
\bra{\bar{\mathcal{G}}}=\bra{\bar{0}}\prod_k \hat{\bar{\eta}}^1_{k-}.
\end{align}
The Loschmidt amplitude at time $t$ is then given by
\begin{align}
G(t)=\bra{\bar{\mathcal{G}}}e^{-i\hat{H}_{QB}^ft}\ket{\mathcal{G}},
\end{align}
where $\hat{H}^f_{QB}$ is the final Hamiltonian.
It can be shown that, for translational symmetric systems, the Loschmidt amplitude factorizes over momentum modes as
\begin{align}
G(t)=\prod_k g_k(t),
\end{align}
with 
\begin{align}\label{g_kt}
g_k(t)=\cos{E^f(k)t}+i{\hat{d}^i(k)\cdot\hat{d}^f(k)}\sin{E^f(k)t},
\end{align}
where $E^f(k)$ are the quasiparticle energies of the final Hamiltonian and $\hat{d}^i(k)$
 and $\hat{d}^f(k)$ are the normalized complex vectors corresponding to initial and final Hamiltonians, respectively, as defined in  Eq. \ref{d_vec}. The detailed derivation of $g_k(t)$ is given in Appendix \ref{der_gk}. Interestingly, this expression is almost same as that of the nSSH2 model except for the normalization factor that is introduced by hand. More importantly, this form remains valid even when pairing terms are present in the initial Hamiltonian too. Using Loschmidt amplitude, the return rate ($RR(t)$) is defined as
\begin{align}\label{RR}
RR(t)=-\frac{1}{N}\log{|G(t)|^2}.
\end{align}
The return rate becomes non-analytic whenever $G(t)=0$, which can happen whenever any of the $g_k(t)=0$, signalling dynamical quantum phase transitions \cite{Heyl2018}. These nonanalyticities correspond to critical times $t_c$ associated with critical momenta $k_c$. If we allow $t$ to be complex, then the complex times $t=i\omega$ at which $g_k(t)=0$ are called Fisher zeroes \cite{Fisher1965}. They are given by
\begin{align}
\omega=i\frac{\pi(2n+1)}{2E^f(k)}+\frac{\tanh^{-1}(\hat{d}^i(k)\cdot\hat{d}^f(k))}{E^f(k)}.
\end{align}
The critical times $t_c$ are determined by the imaginary part of $\omega$:
\begin{align}
t_c^{n\pm}=\frac{\pi(n+\frac{1}{2})\Re[E^f_{k_c^{n\pm}}]+\Im[(E^f_{k_c^{n\pm}})^*\tanh^{-1}(\hat{d}^i_{k_c^{n\pm}}\cdot \hat{d}^f_{k_c^{n\pm}})]}{|E^f_{k_c^{n\pm}}|^2},
\end{align}
where, $k_c^{n\pm}$ is given by the solution of the equation, $\Re[\omega]=0$, i.e.,
\begin{align}
\pi(n+\frac{1}{2})\Im[E^f_{k_c^n}]+\Re[(E^f_{k_c^n})^*\tanh^{-1}(\hat{d}^i_{k_c^{n}}\cdot \hat{d}^f_{k_c^{n}})]=0,\label{kc}
\end{align}
where the $+$ and $-$ in the superscript of $k_c$ are used to denote whether it lies in the plus half or minus half of the BZ. In cases where $E^f(k)$ is real—such as quenches between Hermitian SSH models, $k_c$ is independent of $n$, and the critical times $t_c$ are evenly spaced. However, if $E^f(k)$ is complex, as in quenches involving pairing terms, $k_c$ becomes $n$-dependent, and the critical times are no longer evenly spaced; the intervals between successive $t_c$s vary with $n$ (see Fig. \ref{t_m}).
Lets express $g_k(t)$ as $g_k(t)=r_k(t)e^{i\phi_k(t)}$ with $r_k(t)$ and $\phi_k(t)$ as its magnitude and phase respectively.
A purely geometric and gauge invariant phase, known as the Pancharatnam geometric phase (PGP), $\phi_{\rm pgp}(k,t)$ can be constructed by subtracting the dynamical phase from $\phi_k(t)$.
\begin{align}\label{phi_pgp}
\phi_{\rm pgp}(k,t)=\phi_k(t)-\phi_{dyn}(k,t),
\end{align}
where $\phi_{dyn}(k,t)$ is the dynamical phase given by 
\begin{align}
\phi_{dyn}(k,t)=-i\int_0^tds\bra{\bar{\mathcal{G}}(s)}\frac{d}{ds}\ket{\mathcal{G}(s)},
\end{align}
where $\ket{\mathcal{G}(s)}=e^{-i\hat{H}_{QB}^fs}\ket{\mathcal{G}}$ and $\bra{\bar{\mathcal{G}}(s)}=\bra{\bar{\mathcal{G}}}e^{i\hat{H}_{QB}^fs}$. It can be shown that $\phi_{dyn}(k,t)$ comes out to be 
\begin{align}
\phi_{dyn}(k,t)=E^f(k)\hat{d}^i(k)\cdot\hat{d}^f(k)t.
\end{align}
The DTOP is defined as the winding number of the PGP as a function of momentum at a given time. It can be computed separately over the positive and negative halves of the BZ, denoted by $DTOP_+$ and $DTOP_-$, respectively: 
\begin{align}
DTOP_+(t)&=\frac{1}{2\pi}\int_0^\pi \partial_k \phi_{\rm pgp}(k,t)dk,\notag\\
DTOP_-(t)&=\frac{1}{2\pi}\int_{-\pi}^0 \partial_k \phi_{\rm pgp}(k,t)dk.
\end{align}
\begin{figure}[h]
  \centering
 {\includegraphics[width=0.49\textwidth]{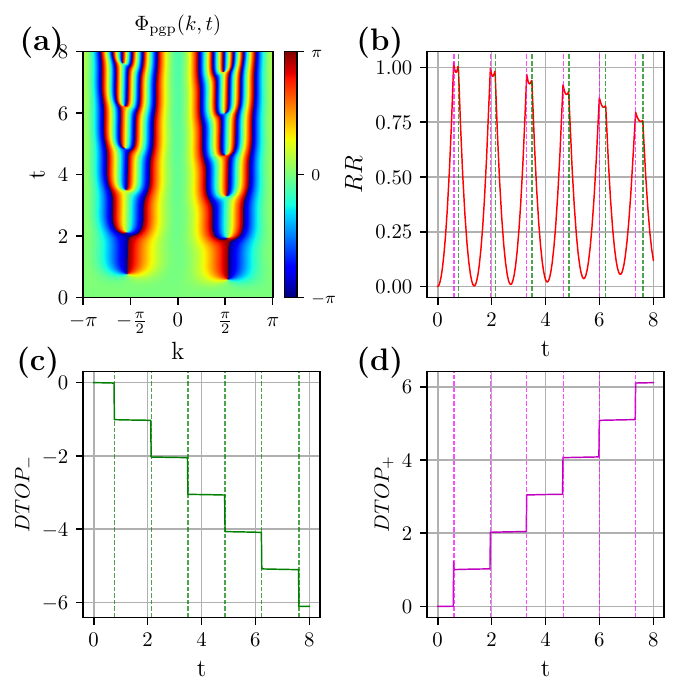}}
  \caption{Dynamical response for the real parameter regime. The Pancharatnam geometric phase, $\phi_{\rm pgp}(k,t)$ in (a), the return rate ($RR(t)$) in (b), the $DTOP_-$ and $DTOP_+$ in (c) and (d), respectively, for the quench from an initial Hamiltonian with parameters $J^i=1$, $\delta^i=-0.9$, $\theta^i=0$ to a final Hamiltonian with parameters $J^f=1$, $\delta^f=0.9$, $\theta^f=0.4$.}
  \label{t_nt}
\end{figure}
In Fig. \ref{t_nt}, we plot the PGP, $\phi_{\rm pgp}(k,t)$ as a function of $k$ and $t$, along with the return rate $RR(t)$ and the DTOPs, for a quench from the initial Hamiltonian in trivial phase ($\nu=0$) with no-pairing term to the final Hamiltonian in non-trivial ($\nu=1$) phase with pairing term. Fig. \ref{t_nt}(a) shows that $\phi_{\rm pgp}(k,t)$ exhibits discontinuities in both halves of the BZ, which leads to quantized jumps in both $DTOP_+$ and $DTOP_-$ as seen in Fig. \ref{t_nt}(c) and (d), respectively. The return rate $RR(t)$, plotted in Fig. \ref{t_nt}(b), exhibits nonanalyticities at the critical times $t_c$, where the Loschmidt amplitude $G(t)$ vanishes. In this setup, the model parameters are parameterized as $v^{i,f}=J^{i,f}(1-\delta^{i,f})$ and $w_r^{i,f}=w_l^{i,f}e^{-\theta^{i,f}}=J^{i,f}(1+\delta^{i,f})$, where superscripts $i$ and $f$ refer to the initial and final Hamiltonians, respectively.

Fig. \ref{t_m} presents a similar analysis for a quench from initial Hamiltonian in trivial phase ($\nu=0$) without pairing term to final Hamiltonian with pairing term in the M{\"o}bius phase ($\nu=1/2$). In contrast to Fig. \ref{t_nt}(a), Fig. \ref{t_m}(a) shows that discontinuities in the PGP appear only in the positive half of the BZ. In the negative half, $\phi_{\rm pgp}(k,t)$ evolves smoothly. Consequently, as shown in  Fig. \ref{t_m}(c) and (d), $DTOP_-=0$ while $DTOP_+$ exhibits quantized jumps. This asymmetry in the dynamical topological response serves as a physical signature of the M{\"o}bius phase, and similar results were reported in Ref. \cite{Nehra2024} for nSSH2 model. The return rate, shown in Fig. \ref{t_m}(b), again becomes non-analytic at critical times due to vanishing of $G(t)$, which in this case arises from critical momentum $k_c$ lying in the positive half of the BZ. While the topological phase transition between different winding numbers is sharply defined in terms of the model's parameters, no such sharp boundary exists following the DTOPs. This is because while the winding numbers are defined by the real part of the Bloch vector, the DTOPs are determined by the complex Bloch vector as in Eq.~\ref{kc}.\\ 
A note about the pseudo-bosonic modes is in order: The vacuum of these modes, as well as the excitations generated by the pseudo-bosonic creation operators, are eigenstates of the Hamiltonian with complex eigenvalues. They are not normalizable and can be understood as quasinormal modes of the inverted harmonic oscillator \cite{Subramanyan2021}, which arise in our system whenever it becomes unstable ($w_r \neq w_l$). Such states belong to the rigged Hilbert space \cite{Ballentine1998}, where the time evolution is non-unitary due to the complex energies, leading to states that either decay or grow with time.

\section{Imaginary Parameter Regime}\label{imag_regime}

We now consider a regime in which the hopping and pairing amplitudes in the Hamiltonian of Eq.~\ref{HQB_realspaceham} are taken to be purely imaginary: $v \rightarrow iv$, $w_r \rightarrow iw_r$, and $w_l \rightarrow iw_l$. We denote the resulting Hamiltonian as $\hat{\tilde{H}}_{QB}$. Although this transformation is simple, it leads to profound changes in the system's dynamical and topological properties, as we show below.

\subsection{Dynamical Matrix}\label{ImDynMat}

In momentum space, the Hamiltonian expressed in the Nambu basis
$
\hat{\phi}_k = (\hat{A}_k, \hat{B}_k, \hat{C}_k, \hat{D}_k, \hat{A}_{-k}^{\dagger}, \hat{B}_{-k}^{\dagger}, \hat{C}_{-k}^{\dagger}, \hat{D}_{-k}^{\dagger})^T
$
takes the form
\begin{align}
\hat{\tilde{H}}_{QB} = \frac{1}{2} \sum_k \hat{\phi}_k^{\dagger} \tilde{H}_{QB}(k) \hat{\phi}_k,
\end{align}
where
\begin{align}
\tilde{H}_{QB}(k) = \begin{pmatrix}
\tilde{P}(k) & \tilde{Q}(k) \\
- \tilde{Q}(k) & - \tilde{P}(k)
\end{pmatrix},
\end{align}
with
\begin{align}
\tilde{P}(k) &= \begin{pmatrix}
0 & if_1(k) & 0 & 0 \\
- if_1^*(k) & 0 & 0 & 0 \\
0 & 0 & 0 & - if_1(k) \\
0 & 0 & if_1^*(k) & 0
\end{pmatrix}, \notag \\
\tilde{Q}(k) &= \begin{pmatrix}
0 & 0 & 0 & if_2(k) \\
0 & 0 & if_2^*(k) & 0 \\
0 & if_2(k) & 0 & 0 \\
if_2^*(k) & 0 & 0 & 0
\end{pmatrix}.
\end{align}

\begin{figure}[h]
  \centering
  {\includegraphics[width=0.49\textwidth]{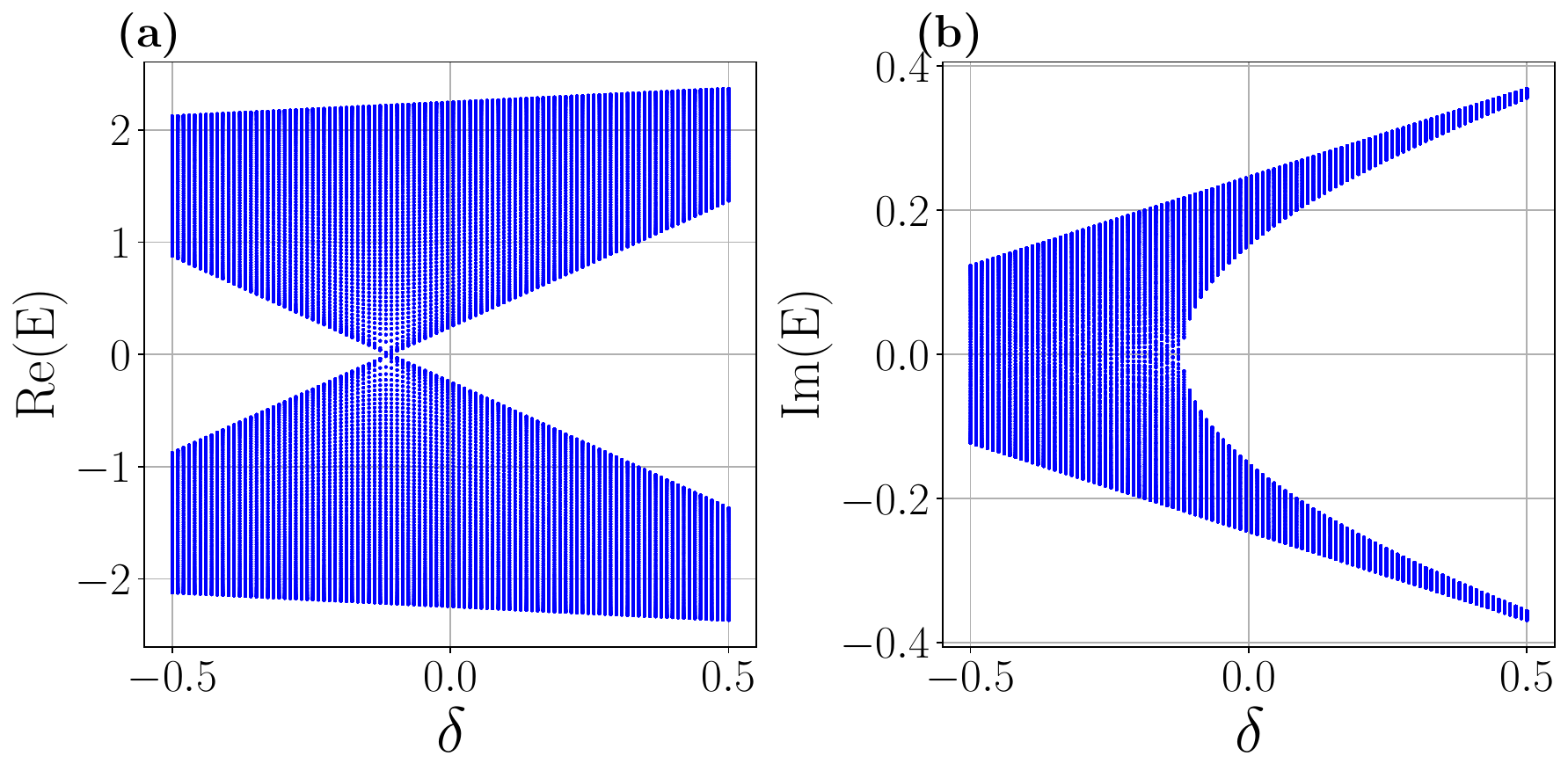}}
  \caption{Real part (a) and imaginary part (b) of the spectrum of $\tilde{G}_{QB}$ in PBC, as function of $\delta$. We fix $J=1$ and $\theta=0.4$ in both the plots.}
  \label{PBC_spec_with_w_imag_hopp}
\end{figure}

The corresponding dynamical matrix is given by $\tilde{G}_{QB}(k) = \tau_3 \tilde{H}_{QB}(k)$. Remarkably, making the hopping and pairing amplitudes imaginary dramatically alters the topological character of the system. We again parameterize as $iv=iJ(1-\delta)$ and $iw_r=iw_le^{-\theta}=iJ(1+\delta)$. To elucidate these changes, we examine the spectrum of $\tilde{G}_{QB}(k)$ as a function of the asymmetry parameter $\delta$, under both PBC and OBC. Fig. \ref{PBC_spec_with_w_imag_hopp} shows the real and imaginary parts of the spectrum in PBC. Compared to the real-parameter case shown in Fig. \ref{PBC_spec_with_w}, several key differences emerge. The M{\"o}bius phase present in $G_{QB}$ is absent in $\tilde{G}_{QB}$. The extended gapless region collapses to a single gap-closing point. For $G_{QB}$, the imaginary part of the spectrum remains gapless across the parameter range, whereas for $\tilde{G}_{QB}$, both the real and imaginary parts become gapped after the transition.

The spectral gap closing point is given by
\begin{align}\label{transition_point}
\delta_0 = \frac{1 - \sqrt{(1 + e^{2\theta}) / 2}}{1 + \sqrt{(1 + e^{2\theta}) / 2}}.
\end{align}
This critical point also signals a topological phase transition, as confirmed by the OBC spectrum in Fig.~\ref{OBC_spec_with_w_imag_hopp}. For $\delta < \delta_0$, the spectrum is gapless in the imaginary part but gapped in the real part. For $\delta > \delta_0$, the spectrum is fully gapped and supports zero modes. We have also checked that there is no NHSE in $\tilde{G}_{QB}$. Another important consequence of introducing imaginary parameters is that the BBC is restored in $\tilde{G}_{QB}$, in contrast to the breakdown observed in $G_{QB}$. To understand this behaviour, we study the symmetries of $\tilde{G}_{QB}$. 

\subsubsection{Symmetries of $\tilde{G}_{QB}$}

By construction, the dynamical matrix $\tilde{G}_{QB}(k)$ exhibits both particle-hole symmetry (PHS1) and pseudo-Hermiticity. These are defined by $\tau_1 \tilde{G}_{QB}^*(-k) \tau_1 = -\tilde{G}_{QB}(k)$ and 
$\tau_3 \tilde{G}_{QB}^\dagger(k) = \tilde{G}_{QB}(k)$, respectively. In addition to PHS1, the matrix $\tilde{G}_{QB}(k)$ possesses two further particle-hole symmetries, PHS3 and PHS4, given by $\Gamma \tilde{G}_{QB}^*(-k) \Gamma^{-1} = -\tilde{G}_{QB}(k)$ and $\tilde{\Gamma} \tilde{G}_{QB}^*(-k) \tilde{\Gamma}^{-1} = -\tilde{G}_{QB}(k)$, respectively,
where $\Gamma = \sigma_x \otimes \sigma_y \otimes \sigma_z$ and $\tilde{\Gamma} = \sigma_z \otimes \sigma_z \otimes \mathbb{I}_2$. Combining PHS3 and PHS4 yields a unitary symmetry $\tilde{U} = \Gamma \tilde{\Gamma}$, under which $\tilde{G}_{QB}(k)$ is invariant: $\tilde{U} \tilde{G}_{QB}(k) \tilde{U}^{-1} = \tilde{G}_{QB}(k)$. This symmetry allows $\tilde{G}_{QB}(k)$ to be block-diagonalized via a unitary transformation. Let $\tilde{Q}$ be the unitary matrix composed of eigenvectors of $\tilde{U}$:
\begin{align}
\tilde{Q} = \frac{1}{\sqrt{2}}\left(
\begin{array}{cccc}
\gamma_1 & 0 & 0 & \gamma_2 \\
0 & \gamma_2 & \gamma_1 & 0 \\
0 & -i\gamma_1 & i\gamma_2 & 0 \\
i\gamma_2 & 0 & 0 & -i\gamma_1 \\
\end{array}
\right),
\end{align}
where,
\begin{align}
\gamma_1 = \begin{pmatrix} i & 0 \\ 0 & 1 \end{pmatrix}, \qquad 
\gamma_2 = \begin{pmatrix} -i & 0 \\ 0 & 1 \end{pmatrix}.
\end{align}

The transformed dynamical matrix $\tilde{G}_b(k)$, defined as
\begin{align}
\tilde{G}_b(k) = \tilde{Q}^\dagger \tilde{G}_{QB}(k) \tilde{Q},
\end{align}
takes a block-diagonal form:
\begin{align}\label{Block_GB1}
\tilde{G}_b(k) = \begin{pmatrix}
H_{nSSH1}(k) & 0 & 0 & 0 \\
0 & -H_{nSSH1}(k) & 0 & 0 \\
0 & 0 & H_{nSSH1}^\dagger(k) & 0 \\
0 & 0 & 0 & -H_{nSSH1}^\dagger(k)
\end{pmatrix},
\end{align}
where each block corresponds to a different version of the nSSH model, say nSSH1, which is given as
\begin{align}
H_{nSSH1}(k) = \begin{pmatrix}
0 & p_1(k) \\
p_2^*(k) & 0
\end{pmatrix},
\end{align}
with
\begin{align}
p_1(k) &= v + \frac{1+i}{2}(w_l - i w_r) e^{-ik}, \\
p_2(k) &= v + \frac{1-i}{2}(w_l + i w_r) e^{-ik}.
\end{align}
Note that this is different than the one in Eq. \ref{HnSSH(k)}.
The block diagonal structure reveals that $\tilde{G}_{QB}(k)$ is unitarily equivalent to a direct sum of four decoupled nSSH1 chains. In the next subsection, we show that $H_{nSSH1}(k)$ does not exhibit the Möbius phase. Moreover, it does not display NHSE, and BBC remains intact despite the system being non-Hermitian. These properties explain the qualitative features observed in the spectrum of $\tilde{G}_{QB}$.

\subsubsection{Topological Phases of the Dynamical Matrix}

Despite the hopping and pairing amplitudes being purely imaginary, the dynamical matrix $\tilde{G}_{QB}(k)$ shows nontrivial topology. This is most transparently understood in terms of its block-diagonal structure. In particular, the topological features of $\tilde{G}_{QB}(k)$ can be characterized through the representative block $H_{nSSH1}(k)$. Analogous to the Hermitian case, this matrix can be expressed as $H_{nSSH1}(k) = \vec{\tilde{d}}(k) \cdot \vec{\sigma}$, where $\vec{\tilde{d}}(k) = \vec{\tilde{d}}^r(k) + i \vec{\tilde{d}}^i(k)$ is a complex Bloch vector, with real and imaginary parts given by:
\begin{align}\label{tilde_d_vec}
\vec{\tilde{d}}^r(k) &= \left(v + \frac{w_l + w_r}{2} \cos k\right) \hat{x} + \frac{w_l + w_r}{2} \sin k \, \hat{y}, \notag \\
\vec{\tilde{d}}^i(k) &= \frac{w_l - w_r}{2} \cos k \, \hat{x} - \frac{w_l - w_r}{2} \sin k \, \hat{y}.
\end{align}
Due to sublattice symmetry, both vectors lie entirely in the $x$-$y$ plane. The eigenvalues of $H_{nSSH1}(k)$ are given by
\begin{align}
\tilde{E}_\pm(k) = \pm \sqrt{p_1(k) p_2^*(k)} = \pm \sqrt{X(k) + iY(k)},
\end{align}
with $X(k) = v^2 + v(w_r + w_l) \cos k + w_r w_l$, and $Y(k) = (w_l - w_r) \bigl(v \cos k + \frac{w_r + w_l}{2}\bigr)$. The EP occurs when both the real and imaginary parts of the square root vanish, i.e., $X(k) = Y(k) = 0$. Solving these equations yields:
\begin{align}
v = \sqrt{\frac{w_r^2 + w_l^2}{2}}, \quad
\text{and} \quad k = \cos^{-1} \left(-\frac{w_r + w_l}{\sqrt{2(w_r^2 + w_l^2)}} \right).
\end{align}
\begin{figure}[h]
  \centering
  {\includegraphics[width=0.49\textwidth]{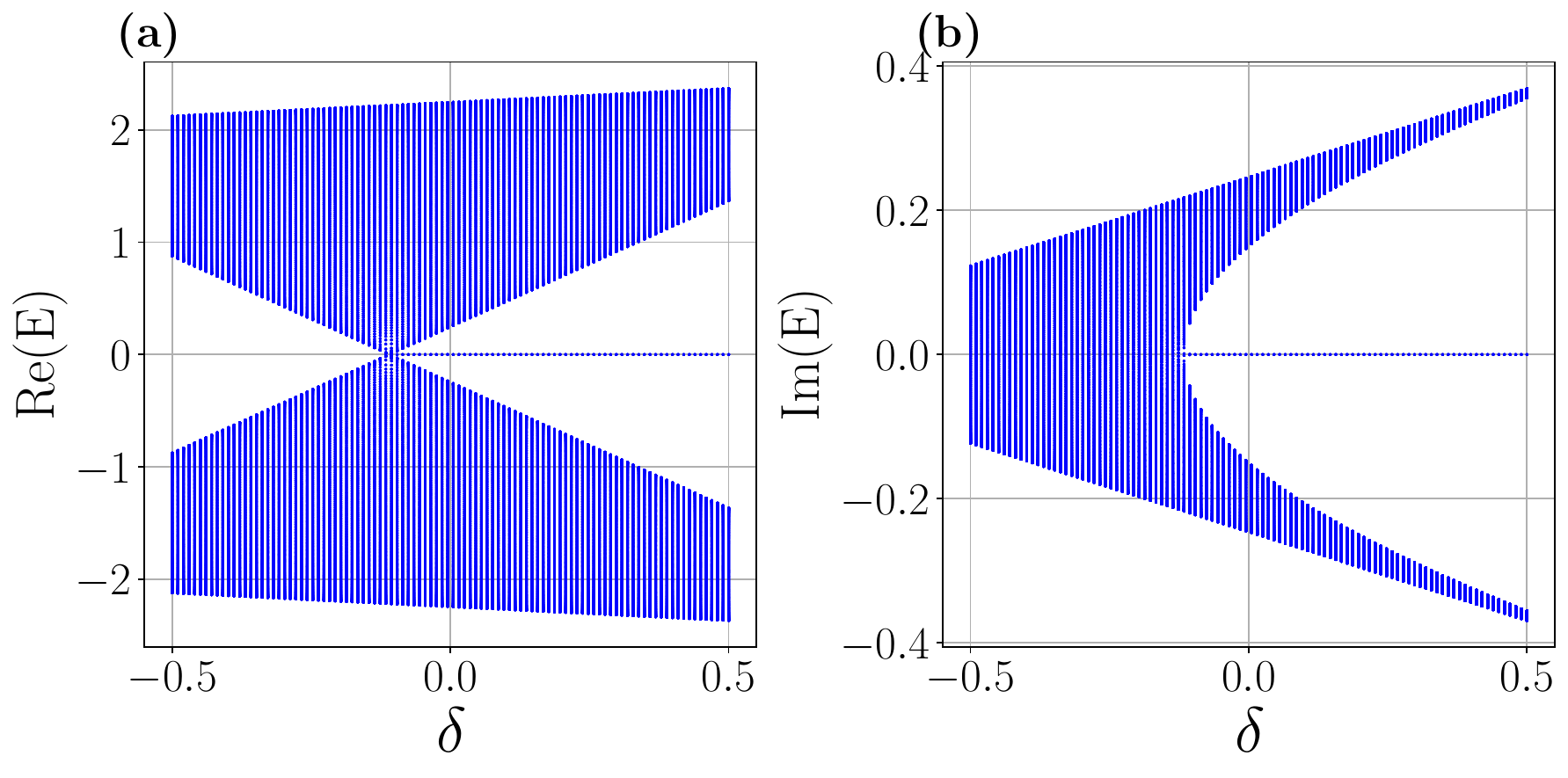}}
  \caption{Real part (a) and imaginary part (b) of the spectrum of $\tilde{G}_{QB}$ in OBC, as function of $\delta$. We fix $J=1$ and $\theta=0.4$ in both the plots.}
  \label{OBC_spec_with_w_imag_hopp}
\end{figure}
Notably, this reduces to the Hermitian transition point $v = w$ when $w_r = w_l = w$. In terms of $J$ and $\delta$, this corresponds to the critical value $\delta_0$ in Eq.~\ref{transition_point}, where $\delta_0 = 0$ in the Hermitian limit.

A key distinction from the real-parameter nSSH2 model is that $H_{nSSH1}(k)$ supports only a single EP, in contrast to the two EPs present in $H_{nSSH2}(k)$. Consequently, the topological classification reduces to two phases: a trivial phase with winding number $\nu = 0$ and a nontrivial phase with $\nu = 1$. The winding number is determined by the closed loop traced out by the real part of the Bloch vector, $\vec{\tilde{d}}^r(k)$, as $k$ varies from $-\pi$ to $\pi$ in the BZ. If this loop encloses the EP, the winding number is $\nu = 1$; otherwise $\nu = 0$. The winding number $\nu = 1$, corresponds to a topologically nontrivial phase with boundary-localized zero modes under OBC (Fig. \ref{OBC_spec_with_w_imag_hopp}), in agreement with the BBC. Importantly, there is no Möbius-like intermediate phase with fractional winding, as is also evident from the PBC spectrum (Fig. \ref{PBC_spec_with_w_imag_hopp}).
\begin{figure*}[t]
  \centering
  {\includegraphics[width=1\textwidth]{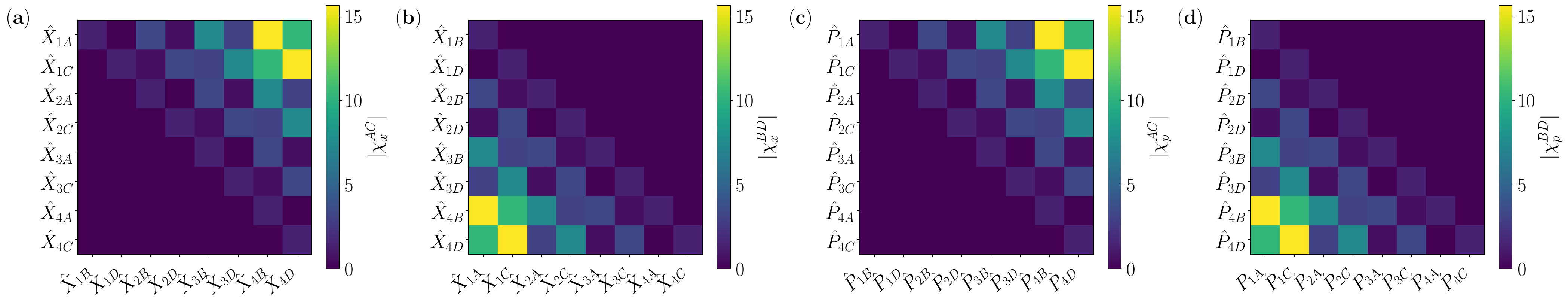}}
  \captionsetup{justification=justified,singlelinecheck=false}
  \caption{Susceptibility matrices showing sublattice-dependent directional amplification. Matrices in (a) $\chi^{AC}_x$ and (c) $\chi^{AC}_p$ show that the quadratures $\hat{X}$ and $\hat{P}$ associated with sublattices $A$ and $C$ are amplified towards left. On the contrast, (b) $\chi^{BD}_x$ and (d) $\chi^{BD}_p$ show that the quadratures $\hat{X}$ and $\hat{P}$ associated with sublattices $B$ and $D$ are amplified towards right.}
  \label{Susc}
\end{figure*}

\subsection{Sublattice-dependent chiral amplification}\label{amplification}
The bosonic Kitaev chain with purely imaginary hopping and pairing terms is known to exhibit quadrature-dependent directional amplification, where the $\hat{X}$ and $\hat{P}$ quadratures amplify in opposite directions~\cite{McDonald2018, Slim2024}. Remarkably, the Hamiltonian $\hat{\tilde{H}}_{QB}$ in Eq.~\ref{HQB_realspaceham}, with purely imaginary hopping and pairing amplitudes, also displays directional amplification. However, in contrast to the bosonic Kitaev chain, the amplification here is not quadrature-dependent but rather sublattice-dependent. To demonstrate this, we express the Hamiltonian in the quadrature basis by writing each bosonic operator as
\begin{align}
\hat{\alpha}_j=\frac{\hat{X}_{j\alpha}+i\hat{P}_{j\alpha}}{\sqrt{2}},
\end{align}
where $\hat{\alpha}_j \in \{\hat{A}_j, \hat{B}_j, \hat{C}_j, \hat{D}_j\}$. The Hamiltonian $\hat{\tilde{H}}_{QB}$ then becomes
\begin{align}\label{Hxp_imag}
&\hat{\tilde{H}}_{QB} = \sum_{j=1}^{N} \biggl[
 -v \big( \hat{X}_{jA} \hat{P}_{jB} - \hat{P}_{jA} \hat{X}_{jB} \big)
 + v \big( \hat{X}_{jC} \hat{P}_{jD} - \hat{P}_{jC} \hat{X}_{jD} \big) \notag\\
&\quad -\frac{w_r+w_l}{2} \Big(
  \hat{X}_{j+1A} \hat{P}_{jB} - \hat{P}_{j+1A} \hat{X}_{jB}
  - \hat{X}_{j+1C} \hat{P}_{jD} + \hat{P}_{j+1C} \hat{X}_{jD} \Big) \notag\\
&\quad +\frac{w_l-w_r}{2} \Big(
  \hat{X}_{jB} \hat{P}_{j+1C} + \hat{X}_{j+1A} \hat{P}_{jD}
  + \hat{P}_{jB} \hat{X}_{j+1C} + \hat{P}_{j+1A} \hat{X}_{jD}
\Big) \biggr].
\end{align}

We now compute the Heisenberg equations of motion for the quadrature operators. Notably, the $\hat{X}$ and $\hat{P}$ quadratures evolve independently. For $\hat{X}$ quadratures, we have:
\begin{align}\label{Heis_X}
\dot{\hat{X}}_{jA}&=v\hat{X}_{jB}+\frac{w_r+w_l}{2}\hat{X}_{j-1B}+\frac{w_l-w_r}{2}\hat{X}_{j-1D}-\frac{\gamma}{2}\hat{X}_{jA}-\sqrt{\gamma}\hat{X}_{jA}^{IN},\notag\\
\dot{\hat{X}}_{jB}&=-v\hat{X}_{jA}-\frac{w_r+w_l}{2}\hat{X}_{j+1A}+\frac{w_l-w_r}{2}\hat{X}_{j+1C}-\frac{\gamma}{2}\hat{X}_{jB}-\sqrt{\gamma}\hat{X}_{jB}^{IN},\notag\\
\dot{\hat{X}}_{jC}&=-v\hat{X}_{jD}-\frac{w_r+w_l}{2}\hat{X}_{j-1D}+\frac{w_l-w_r}{2}\hat{X}_{j-1B}-\frac{\gamma}{2}\hat{X}_{jC}-\sqrt{\gamma}\hat{X}_{jC}^{IN},\notag\\
\dot{\hat{X}}_{jD}&=v\hat{X}_{jC}+\frac{w_r+w_l}{2}\hat{X}_{j+1C}+\frac{w_l-w_r}{2}\hat{X}_{j+1A}-\frac{\gamma}{2}\hat{X}_{jD}-\sqrt{\gamma}\hat{X}_{jD}^{IN},\notag\\
\end{align}
and for the $\hat{P}$ quadratures:
\begin{align}\label{Heis_P}
\dot{\hat{P}}_{jA}&=v\hat{P}_{jB}+\frac{w_r+w_l}{2}\hat{P}_{j-1B}-\frac{w_l-w_r}{2}\hat{P}_{j-1D}-\frac{\gamma}{2}\hat{P}_{jA}-\sqrt{\gamma}\hat{P}_{jA}^{IN},\notag\\
\dot{\hat{P}}_{jB}&=-v\hat{P}_{jA}-\frac{w_r+w_l}{2}\hat{P}_{j+1A}-\frac{w_l-w_r}{2}\hat{P}_{j+1C}-\frac{\gamma}{2}\hat{P}_{jB}-\sqrt{\gamma}\hat{P}_{jB}^{IN},\notag\\
\dot{\hat{P}}_{jC}&=-v\hat{P}_{jD}-\frac{w_r+w_l}{2}\hat{P}_{j-1D}-\frac{w_l-w_r}{2}\hat{P}_{j-1B}-\frac{\gamma}{2}\hat{P}_{jC}-\sqrt{\gamma}\hat{P}_{jC}^{IN},\notag\\
\dot{\hat{P}}_{jD}&=v\hat{P}_{jC}+\frac{w_r+w_l}{2}\hat{P}_{j+1C}-\frac{w_l-w_r}{2}\hat{P}_{j+1A}-\frac{\gamma}{2}\hat{P}_{jD}-\sqrt{\gamma}\hat{P}_{jD}^{IN},
\end{align}
where $\gamma$ denotes the damping or dissipation rates. These equations can be written in the matrix form as
\begin{align}\label{xp_mat}
\begin{pmatrix}
\dot{\hat{\textbf{X}}} \\
\dot{\hat{\textbf{P}}} 
\end{pmatrix}&=\begin{pmatrix}
\textbf{h}_x-\frac{\gamma}{2}\textbf{1} & \textbf{0} \\
\textbf{0} & \textbf{h}_p-\frac{\gamma}{2}\textbf{1}
\end{pmatrix}\begin{pmatrix}
\hat{\textbf{X}} \\
\hat{\textbf{P}}
\end{pmatrix}-\sqrt{\gamma}\begin{pmatrix}
\hat{\textbf{X}}^{IN} \\
\hat{\textbf{P}}^{IN}
\end{pmatrix}\notag\\
&=\textbf{M}\begin{pmatrix}
\hat{\textbf{X}} \\
\hat{\textbf{P}}
\end{pmatrix}-\sqrt{\gamma}\begin{pmatrix}
\hat{\textbf{X}}^{IN} \\
\hat{\textbf{P}}^{IN}
\end{pmatrix},
\end{align}
where $\hat{\textbf{X}}= (\hat{X}_{1A}, \hat{X}_{1B}, \hat{X}_{1C}, \hat{X}_{1D},...,\hat{X}_{NA}, \hat{X}_{NB}, \hat{X}_{NC}, \hat{X}_{ND})^T$ and similarly for $\hat{P}$. The vectors $\hat{\textbf{X}}^{IN}$ and $\hat{\textbf{P}}^{IN}$ are the corresponding input modes associated with $\hat{X}$ and $\hat{P}$ quadratures. The matrices $\textbf{h}_x$ and $\textbf{h}_p$ are the non-Hermitian dynamical matrices corresponding to $\hat{X}$ and $\hat{P}$ quadratures (see App. \ref{App.susc}), and $M$ denotes the full dynamical matrix in quadrature basis.

In small dissipation limit ($\gamma \rightarrow 0$), the $k$-space dynamical matrix $M(k)$ is given by
\begin{align}
 M(k)&=\begin{pmatrix}\textbf{h}_x(k)&\textbf{0}\\\textbf{0}&\textbf{h}_p(k)\end{pmatrix},
\end{align}
where 
\begin{align}
\textbf{h}_x(k)&=\begin{pmatrix}0&f_1(k)&0&f_2(k)\\-f^*_1(k)&0&f^*_2(k)&0\\0&f_2(k)&0&-f_1(k)\\f^*_2(k)&0&f^*_1(k)&0\end{pmatrix},\notag\\
 \textbf{h}_p(k)&=\begin{pmatrix}0&f^*_1(k)&0&-f^*_2(k)\\-f_1(k)&0&-f_2(k)&0\\0&-f^*_2(k)&0&-f^*_1(k)\\-f_2(k)&0&f_1(k)&0\end{pmatrix}.
\end{align}
To understand sublattice-dependent chiral amplification, we take $\gamma \rightarrow 0$ and we analyze how the $\hat{X}$ and $\hat{P}$ quadratures evolve across different sublattices. The effect of finite $\gamma$ is discussed in App.~\ref{App.susc}. While the $\hat{X}$ and $\hat{P}$ quadratures are decoupled from each other, the sublattice degrees of freedom associated with each quadrature remain coupled. It is not immediately clear whether directional amplification occurs, or which sublattices amplify in which direction.

To investigate this, we compute the susceptibility matrix $\chi$ under OBC, defined as the inverse of the dynamical matrix $M$ \cite{Slim2024}. Since $M$ is block-diagonal, $\chi$ also decomposes into two blocks: $\chi_x = \textbf{h}_x^{-1}$ and $\chi_p = \textbf{h}_p^{-1}$, corresponding to the $\hat{X}$ and $\hat{P}$ quadratures, respectively.

Each quadrature couples only specific sublattices: sublattices $A$ and $C$ both couple only to sublattices $B$ and $D$, and vice versa. To capture this structure, we reorganize $\chi_x$ and $\chi_p$ into sub-matrices. Specifically, we define $\chi^{AC}_{x,p}$ as the sub-matrices with rows corresponding to $A$ and $C$ and columns to $B$ and $D$, and $\chi^{BD}_{x,p}$ as the sub-matrices with rows for $B$ and $D$ and columns for $A$ and $C$ (see Fig. \ref{Susc}).

Fig. \ref{Susc} shows the heatmaps of these sub-matrices. Figs. \ref{Susc}(a,c) clearly show that the $\hat{X}$ and $\hat{P}$ quadratures on sublattices $A$ and $C$ experience amplification towards the left, while Figs. \ref{Susc}(b,d) reveal that the sublattices $B$ and $D$ are amplified towards the right. Thus, the amplification is not quadrature-dependent (as in earlier work \cite{McDonald2018, Slim2024}), but sublattice-dependent.

This sublattice-dependent chiral amplification has topological origin \cite{Slim2024}. To illustrate this, we consider the symmetric limit $w_r=w_l=w$ (or equivalently, $\theta=0$), in which the topological phase transition occurs at $\delta_0=0$ (i.e., $v=w$). The system is topologically trivial for $v>w$ and non-trivial for $v<w$. In this limit, the susceptibility matrices for a small system ($N=4$) take simple form:
\[
|\chi^{AC}_{x,p}|=\frac{1}{v}\begin{pmatrix}
1 & 0 & G_0 & 0 & G_0^2 & 0 & G_0^3 & 0 \\
0 & 1 & 0 & G_0 & 0 & G_0^2 & 0 & G_0^3 \\
0 & 0 & 1 & 0 & G_0 & 0 & G_0^2 & 0 \\
0 & 0 & 0 & 1 & 0 & G_0 & 0 & G_0^2 \\
0 & 0 & 0 & 0 & 1 & 0 & G_0 & 0 \\
0 & 0 & 0 & 0 & 0 & 1 & 0 & G_0 \\
0 & 0 & 0 & 0 & 0 & 0 & 1 & 0 \\
0 & 0 & 0 & 0 & 0 & 0 & 0 & 1
\end{pmatrix},
\]
and
\[
|\chi^{BD}_{x,p}|=\frac{1}{v}\begin{pmatrix}
1 & 0 & 0 & 0 & 0 & 0 & 0 & 0 \\
0 & 1 & 0 & 0 & 0 & 0 & 0 & 0 \\
G_0 & 0 & 1 & 0 & 0 & 0 & 0 & 0 \\
0 & G_0 & 0 & 1 & 0 & 0 & 0 & 0 \\
G_0^2 & 0 & G_0 & 0 & 1 & 0 & 0 & 0 \\
0 & G_0^2 & 0 & G_0 & 0 & 1 & 0 & 0 \\
G_0^3 & 0 & G_0^2 & 0 & G_0 & 0 & 1 & 0 \\
0 & G_0^3 & 0 & G_0^2 & 0 & G_0 & 0 & 1
\end{pmatrix},
\]
where $G_0=w/v$, and the modulus signs denote that we are taking absolute value of each of the matrix element. For $\chi^{AC}_{x}$, the basis is ordered such that the rows correspond to $(\hat{X}_{1A}, \hat{X}_{1C}, \hat{X}_{2A}, \hat{X}_{2C},...,\hat{X}_{NA}, \hat{X}_{NC})$, while the columns correspond to $(\hat{X}_{1B}, \hat{X}_{1D}, \hat{X}_{2B}, \hat{X}_{2D},...,\hat{X}_{NB}, \hat{X}_{ND})$ and similarly for $\chi^{AC}_{p}$. For $\chi^{BD}_{x}$, the ordering is reversed: the rows correspond to $(\hat{X}_{1B}, \hat{X}_{1D}, \hat{X}_{2B}, \hat{X}_{2D},...,\hat{X}_{NB}, \hat{X}_{ND})$, while the columns correspond to $(\hat{X}_{1A}, \hat{X}_{1C}, \hat{X}_{2A}, \hat{X}_{2C},...,\hat{X}_{NA}, \hat{X}_{NC})$ and similarly for $\chi^{BD}_{p}$. These matrices demonstrate that directional amplification emerges only in the topologically non-trivial regime $G_0>1$, establishing a clear link between topology and amplification. This connection with non-trivial topology survives even when $w_r\neq w_l$ (i.e., $\theta\neq 0$). In this case too, the amplification occurs only in the non-trivial region, i.e., $\delta>\delta_0$ given by Eq. \ref{transition_point}.

We note that changing the hopping and pairing strengths in the QBH from real to purely imaginary leads to significant changes in both the energy spectrum and the topology of the dynamical matrix. We observe a similar behavior in the bosonic Kitaev chain: when its hopping and pairing terms are purely imaginary, the dynamical matrix becomes unitarily equivalent to the Hatano–Nelson model \cite{Hatano1997}, as shown in \cite{McDonald2018, Slim2024}. However, when these terms are taken to be real, this equivalence no longer holds. In fact, the dynamical matrix in that case exhibits a purely real spectrum—even under PBC.

\section{Physical Realization} \label{Physical_realization}

The QBH model discussed in this work can be realized in a variety of bosonic platforms such as optomechanical arrays and superconducting quantum circuits. Recently, the bosonic Kitaev chain has been experimentally realized in a nano-optomechanical network \cite{Slim2024} and in synthetic dimensions of a multimode superconducting parametric cavity \cite{Busnaina2024}. Since our proposed model can be viewed as a generalized bosonic Kitaev chain, these experimental realizations can readily be extended to implement our model. The effective hopping and pairing interactions between bosonic modes in these experiments are generated by parametric drivings or temporal modulation of the driving beams. Notably, the temporal modulation of the driving or pump beams at the frequency difference $|\omega_i-\omega_j|$ (or sum $|\omega_i+\omega_j|$) generates a complex hopping (or pairing) term between $i$ and $j$ bosonic modes with bare frequency $\omega_i$, $\omega_j$ via a beam-splitter (two-mode squeezing) interaction. The magnitudes and phases of the complex hopping and pairing terms are determined by the magnitudes and phases of these driving beams, which can be precisely controlled in experiments \cite{Slim2024, Busnaina2024, Dong2015}. By combining and phase-controlling such drivings, one can synthesize the full QBH explored in this work. 

In particular, the distinction between the real and imaginary parameter regimes in our model corresponds to different choices of the phases of the external drives that set the phases of the complex hopping and pairing terms. When the driving phases are tuned such that the hopping and pairing terms are real, the system realizes the real parameter regime exhibiting chiral topological dynamics. One can similarly tune the driving phases to make the hopping and pairing terms purely imaginary, thereby realizing the imaginary parameter regime associated with topological amplification. Thus, both these regimes can be implemented within the same physical setup by controlling only the phase of the external driving or pump beams.


\section{Summary and outlook}\label{sum}
In this work, we introduced a Hermitian quadratic bosonic model, whose dynamical matrix exhibits features of non-reciprocal hopping SSH chains. We have shown that by tuning the hopping and pairing amplitudes from real to purely imaginary values, one can access distinct dynamical behaviors and topological phases—ranging from chiral dynamical topological response in the real regime to directional amplification tied to non-trivial topology in the imaginary regime. 

The dynamical matrix is unitarily equivalent to four decoupled copies of the nSSH2 model in the real parameter regime. This equivalence allowed us to identify the topological phases of the dynamical matrix, including the M{\"o}bius phase—a gapless topological phase with a fractional winding number and no counterpart in Hermitian systems. We demonstrated that QBH Hamiltonian reproduces the dynamics of the nSSH2 model, including its Loschmidt amplitude and Pancharatnam geometric phase. By computing the DTOP, we revealed a chiral dynamical response unique to the M{\"o}bius phase, distinct from those in the trivial and non-trivial phases. We note that the dynamical matrix has complex eigenvalues when $w_r\neq w_l$. When written in a quadrature basis, our Hamiltonian can be interpreted as a collection of coupled oscillators, which become inverted when $w_r\neq w_l$. The states we use to calculate the Loschmidt amplitude can be interpreted as manifestations of quasi-normal modes of these inverted oscillators since they are the eigenstates with complex eigenvalues \cite{Subramanyan2021}.

In the imaginary parameter regime, the dynamical matrix becomes unitarily equivalent to a different non-Hermitian SSH variant (nSSH1), which hosts only two topological phases—trivial and non-trivial. We showed that the QBH in this regime exhibits sublattice-dependent chiral amplification under OBC. This directional amplification originates from the non-trivial topology of the dynamical matrix and highlights the subtle and rich connection between topology and directional amplification.

We emphasize that the phenomena discussed in this work—namely, the emergence of effective non-Hermitian dynamics, amplification, and Möbius-type topology—are unique to bosonic quadratic Hamiltonians. In these systems, the canonical commutation relations impose constraints on the diagonalization that render the associated dynamical matrix non-Hermitian even though the underlying Hamiltonian is Hermitian (see App. \ref{NH_dyn_frm_Herm_QBH}). This diagonalization
 procedure (by a non-unitary Bogoliubov transformation) enables the emergence of non-Hermitian topology, amplification, and non-trivial chiral responses in the DTOP. In contrast, for fermionic quadratic Hamiltonians, the diagonalization is governed by unitary transformations that preserve Hermiticity of both the Hamiltonian and the dynamical matrix. Consequently, fermionic systems do not exhibit such non-Hermitian-like features. 

Our results demonstrate that a fully Hermitian bosonic Hamiltonian can naturally encode non-Hermitian topology and dynamics through its dynamical matrix. This provides a consistent and physically motivated framework without the ambiguities inherent in non-Hermitian quantum mechanics. This framework offers a promising path for realizing non-Hermitian phenomena in experimentally accessible Hermitian platforms, such as superconducting circuits and optomechanical systems with engineered quadratic couplings. The interplay between real and imaginary parameter regimes within a single Hermitian system suggests new ways to probe and control topological phases dynamically. Related approaches to probing bulk topology, such as extracting winding numbers from spectral functions via ARPES or STM \cite{Estake2025}, may offer complementary experimental perspectives. Recent work on topological scattering resonances at ultralow frequencies \cite{Hassani2020} likewise illustrates how topology enables unconventional and robust responses across different physical platforms.

\section{Acknowledgements}
We thank Arul Lakshminarayan for valuable discussions. K.B.E. acknowledges Vijay Kumar and Rupak Bag for helpful discussions.

\appendix
\section{Derivation of winding number formula}\label{App.WN}
In this appendix, we derive the winding number formula for sublattice symmetric non-Hermitian 1D systems. The discussion in this section is following \cite{Yin2018}. We start with a generic sublattice-symmetric Hamiltonian matrix $H(k)=\vec{d}(k)\cdot \vec{\sigma}$. The energies for this Hamiltonian are given by $E_{\pm}(k)=\pm\sqrt{d_x^2(k)+d_y^2(k)}$, where $d_x(k)=d_x^r(k)+id_x^i(k)$ and $d_y(k)=d_y^r(k)+id_y^i(k)$ are complex $x$ and $y$ components of the Bloch vector $\vec{d}(k)$. The superscripts $r$ and $i$ denote the real and imaginary parts. The EPs are given by 
\begin{align}\label{EP_conditions}
EP1&: d_x^r(k)=d_y^i(k), \quad d_y^r(k)=-d_x^i(k),\notag\\
EP2&: d_x^r(k)=-d_y^i(k), \quad d_y^r(k)=d_x^i(k). 
\end{align}
Generic Hamiltonians of such type have two EPs like the nSSH2 has. Sometimes both these conditions in Eq. \ref{EP_conditions} can give the same EP, which happens for the nSSH1. The winding number, which serves as topological invariant, is defined using $\vec{d}(k)$ as
\begin{align}\label{WN1}
\nu=\frac{1}{2\pi}\int_0^{2\pi}dk\frac{d_x(k)\partial_k d_y(k)-d_y(k)\partial_k d_x(k)}{d_x^2(k)+d_y^2(k)}.
\end{align}
Introducing a complex angle $\phi(k)$ via $\tan \phi(k)=d_y(k)/d_x(k)$, the above expression can be rewritten as
\begin{align}\label{WNphik}
\nu=\frac{1}{2\pi}\oint_c dk\frac{\partial_k \phi(k)}{dk},
\end{align}
where the contour $c$ corresponds to $k\in [0,2\pi]$. Here, $\phi(k)$ is well defined everywhere except at the EPs and it can also be written as 
\begin{align}
e^{i2\phi(k)}=\frac{d_x(k)+id_y(k)}{d_x(k)-id_y(k)}.
\end{align}
Since $\phi(k)$ is a complex angle, it can be expressed as 
\begin{align}
\phi(k)=\phi_r(k)+i\phi_i(k),
\end{align}
where $\phi_r(k)$ and $\phi_i(k)$ denote its real and imaginary parts, respectively. It can be checked that 
\begin{align}
e^{-2\phi_i(k)}=\abs{\frac{d_x(k)+id_y(k)}{d_x(k)-id_y(k)}},
\end{align}
which is a continuous periodic function of $k$. So,
\begin{align}\label{phi_ik}
\oint_c dk\frac{\partial_k \phi_i(k)}{dk}=\phi_i(2\pi)-\phi_i(0)=0.
\end{align}
This means that the winding number has no contribution from the imaginary part of $\phi(k)$ and is entirely determined by $\phi_r(k)$. $\phi_r(k)$ is given by
\begin{align}
e^{i2\phi_r(k)}=\left. \frac{d_x(k)+id_y(k)}{d_x(k)-id_y(k)} \right/ \abs{\frac{d_x(k)+id_y(k)}{d_x(k)-id_y(k)}}, 
\end{align}
which gives 
\begin{align}
\tan 2\phi_r(k)=\left. \Im \frac{d_x(k)+id_y(k)}{d_x(k)-id_y(k)} \right/ \Re \frac{d_x(k)+id_y(k)}{d_x(k)-id_y(k)}.
\end{align}
It can be shown after doing some algebra that 
\begin{align}\label{twophi}
\tan 2\phi_r(k)=\tan(\phi_1(k)+\phi_2(k)),
\end{align}
where
\begin{align}\label{tanphi1phi2}
\tan \phi_1(k)=\frac{d_y^r(k)+d_x^i(k)}{d_x^r(k)-d_y^i(k)}, \quad \tan \phi_2(k)=\frac{d_y^r(k)-d_x^i(k)}{d_x^r(k)+d_y^i(k)}.
\end{align}
The angles $\phi_{1,2}(k)$ can be understood as the angles subtended by the real part of $\vec{d}(k)$ at the two EPs given by Eq. \ref{EP_conditions}. This can be understood by shifting the origin of the $d_x^r(k)$-$d_y^r(k)$ plane to EP1 and EP2, which gives $\tan \phi_1(k)$ and $\tan \phi_2(k)$ to be exactly given by Eq. \ref{tanphi1phi2}. Eq. \ref{twophi} implies $\phi_r(k)=(\phi_1(k)+\phi_2(k))/2+n\pi,$ where $n$ is an integer. Substituting this in Eq. \ref{WNphik} and using the Eq. \ref{phi_ik}, we get
\begin{align}\label{total_nu}
\nu=\frac{\nu_1+\nu_2}{2},
\end{align}
where
\begin{align}
\nu_{1,2}&=\frac{1}{2\pi}\oint_c dk\frac{\partial_k \phi_{1,2}(k)}{dk}.
\end{align}
Therefore, the winding number $\nu$ can be interpreted as the average of two winding numbers, $\nu_1$ and $\nu_2$, each associated with an EP.\\

\section{Effective non-Hermitian dynamics from quadratic bosonic Hamiltonian}\label{NH_dyn_frm_Herm_QBH}
We briefly discuss how non-Hermitian dynamics emerges in quadratic bosonic Hamiltonians without total particle number conservation. Some of this discussion is following \cite{Flynn2020}. Let us consider a general form of the QBH,

\begin{align}\label{H_herm}
\hat{H}&= \sum_{i,j =1}^{N} K_{ij}\hat{a}_{i}^{\dagger} \hat{a}_{j} + \frac{1}{2}\bigl(\Delta_{ij}\hat{a}^{\dagger}_i \hat{a}_j^\dagger+\Delta^*_{ij}\hat{a}_i\hat{a}_j\bigr) \notag\\
	&=\frac{1}{2}\hat{\Phi}^\dagger H{\hat\Phi}-\frac{1}{2} \tr K,
\end{align}

where $\hat\Phi=(\hat{a}_1,...,\hat{a}_N,\hat{a}_1^\dagger,..,\hat{a}_N^\dagger)^T$ is the Nambu array composed of bosonic annihilation and creation operators $\hat{a}_i$ and $\hat{a}_i^\dagger$ at site $i$. $H$ is the Hamiltonian matrix which is given by 

\begin{align}
H&=\begin{pmatrix}K&\Delta\\\Delta^*&K^T\end{pmatrix}.
\end{align}

The Hermiticity of $\hat{H}$ implies Hermiticity of $K$ i.e. $K^T=K^*$. The bosonic commutation relations imply $\Delta^T=\Delta$. The two relations together imply $H^\dagger=H$ for the matrix $H$. It also satisfies the following relation

\begin{align}
H^*=\Gamma_1H\Gamma_1, 
\end{align} 

where, $\Gamma_i=\sigma_i\otimes \mathbb{I}_{N}$ with $\mathbb{I}_{N}$ being the $N$ dimensional identity matrix. Using the Heisenberg equations for $a_i$ and $a_i^\dagger$, we can write down the equation for the  Nambu array $\hat{\Phi}$. It turns out to be \cite{Flynn2020} (we take $\hbar=1$)

\begin{align}
i\frac{d}{dt}\hat{\Phi}(t)=G\hat{\Phi}(t), \quad G=\Gamma_3 H=\begin{pmatrix}K&\Delta\\-\Delta^*&-K^T\end{pmatrix}.
\end{align} 

The matrix $G$ is known as the dynamical matrix. $G$ is a non-Hermitian matrix since
\begin{align}
G^\dagger=H\Gamma_3=\begin{pmatrix}K&-\Delta\\\Delta^*&-K^T\end{pmatrix}\neq G.
\end{align} 
Therefore the eigenvalues of $G$ are complex in general \cite{Flynn2020}.\\

On the other hand, it can be verified that if the Hamiltonian in Eq. \ref{H_herm} is fermionic, it implies that $\Delta^T=-\Delta$ and $K$ remains Hermitian, $K^T=K^*$. And if we calculate the dynamical matrix for fermions, it turns out to be 
\begin{align}
G=\begin{pmatrix}K&\Delta\\-\Delta^*&-K^T\end{pmatrix} \quad \Rightarrow \quad G^\dagger=\begin{pmatrix}K&\Delta\\-\Delta^*&-K^T\end{pmatrix}= G,
\end{align}
where we have used $K^T=K^*$ and $\Delta^T=-\Delta$. So, the dynamical matrix becomes non-Hermitian due to the bosonic statistics and the presence of particle non-conserving terms in the Hamiltonian.\\

\section{Derivation of $g_k(t)$}\label{der_gk}
In this Appendix, we derive the expression for $g_k(t)$ in Eq. \ref{g_kt}. The $k$-space Loschmidt amplitude $g_k(t)$ is defined as
\begin{align}
g_k(t)=\bra{\bar{k}_-^i}e^{-i\hat{H}_{QB}^ft}\ket{k_-^i},
\end{align}
where $\ket{k_-^i}$ single particle eigenstate of the initial Hamiltonian with eigenvalue $-E^i(k)$, and $\bra{\bar{k}_-^i}$ is its biorthogonal counterpart. They are given by
\begin{align}
\ket{k_-^i}=\hat{\eta}_{k-}^{1\dagger} \ket{0},
\end{align}
and 
\begin{align}
\bra{\bar{k}_-^i}=\bra{\bar{0}}\hat{\bar{\eta}}^1_{k-}.
\end{align}
To analytically calculate $e^{-i\hat{H}_{QB}^ft}\ket{k_-^i}$, we need to write final Hamiltonian $\hat{H}_{QB}^f$ in its normal form as in Eq. \ref{normal_Hf} by doing spectral decomposition of the dynamical matrix. The final Hamiltonian in normal form is given by
\begin{align}
\hat{H}^f_{QB} &= \sum_{k}E^f(k){\hat{\zeta}_{k+}^{1\dagger}} \hat{\bar{\zeta}}^1_{k+}+E^f(k){\hat{\zeta}_{k+}^{2\dagger}} \hat{\bar{\zeta}}^{2}_{k+}\notag\\
&-E^f(k){\hat{\zeta}_{k-}^{1\dagger}} \hat{\bar{\zeta}}^1_{k-}-E^f(k){\hat{\zeta}_{k-}^{2\dagger}} \hat{\bar{\zeta}}^{2}_{k-},
\end{align}
where $\hat{\zeta}_{k\pm}^{1,2\dagger}$ and $\hat{\bar{\zeta}}_{k\pm}^{1,2}$ are the pseudo-bosonic normal mode creation and annihilation operators associated with the final Hamiltonian. By working out the relation between the initial and final normal modes, it can be checked that the vacuum of the final pseudo-bosonic modes is same as that of the initial one. Here, we give the relation of $\hat{\eta}_{k-}^{1\dagger}$ and $\hat{\bar{\eta}}^1_{k-}$ with the final normal mode operators.
\begin{align}
\hat{\eta}_{k-}^{1\dagger}&=\frac{F_2(k)}{F_2^2(k)-F_1^2(k)}\hat{\zeta}_{k+}^{1\dagger}-\frac{F_1(k)}{F_2^2(k)-F_1^2(k)}\hat{\zeta}_{k-}^{1\dagger}\notag\\
\hat{\bar{\eta}}^1_{k-}&=\frac{Q_2(k)}{Q_2^2(k)-Q_1^2(k)}\hat{\bar{\zeta}}^1_{k+}-\frac{Q_1(k)}{Q_2^2(k)-Q_1^2(k)}\hat{\bar{\zeta}}^1_{k-},
\end{align}
where
\begin{align}
F_{1,2}&=\frac{1}{2}\Bigl(1\pm \sqrt{\frac{f^f_1(k)f^{i*}_2(k)}{f^{f*}_2(k)f^{i}_1(k)}}\Bigr),\notag\\
Q_{1,2}&=\frac{1}{2}\Bigl(1\pm \sqrt{\frac{f^{f*}_2(k)f^{i}_1(k)}{f^{f}_1(k)f^{i*}_2(k)}}\Bigr),
\end{align}
where $f_{1,2}(k)$ are the complex functions given in Eq. \ref{f1f2} and the superscripts $i$ and $f$ denote whether they are associated with initial or final Hamiltonian. We use this to calculate $e^{-i\hat{H}_{QB}^ft}\ket{k_-^i}$.
\begin{align}
e^{-i\hat{H}_{QB}^ft}\ket{k_-^i}&=e^{-i\hat{H}_{QB}^ft}\Bigl(\frac{F_2(k)}{F_2^2(k)-F_1^2(k)}\ket{k_+^f}\notag\\
&-\frac{F_1(k)}{F_2^2(k)-F_1^2(k)}\ket{k_-^f}\Bigr),
\end{align}
where $\ket{k_\pm^f}=\hat{\zeta}_{k\pm}^{1\dagger}\ket{0}.$ This gives
\begin{align}
e^{-i\hat{H}_{QB}^ft}\ket{k_-^i}&=\frac{F_2(k)e^{-iE^f(k)t}}{F_2^2(k)-F_1^2(k)}\ket{k_+^f}-\frac{F_1(k)e^{iE^f(k)t}}{F_2^2(k)-F_1^2(k)}\ket{k_-^f}.
\end{align}
Similarly, we can express $\bra{\bar{k}_-^i}$ as
\begin{align}
\bra{\bar{k}_-^i}=\bra{\bar{k}_+^f}\frac{Q_2(k)}{Q_2^2(k)-Q_1^2(k)}-\bra{\bar{k}_-^f}\frac{Q_1(k)}{Q_2^2(k)-Q_1^2(k)},
\end{align}
where $\bra{\bar{k}_\pm^f}=\bra{\bar{0}}\hat{\bar{\zeta}}^1_{k\pm}.$ Finally, by taking inner product of this with $e^{-i\hat{H}_{QB}^ft}\ket{k_-^i}$, we get
\begin{align}
g_k(t)&=\bra{\bar{k}_-^i}e^{-i\hat{H}_{QB}^ft}\ket{k_-^i}\notag\\
&=\frac{Q_2(k)F_2(k)e^{-iE^f(k)t}+Q_1(k)F_1(k)e^{iE^f(k)t}}{(Q_2^2(k)-Q_1^2(k))(F_2^2(k)-F_1^2(k))}.
\end{align}
It turns out that, $(Q_2^2(k)-Q_1^2(k))(F_2^2(k)-F_1^2(k))=1$ and 
\begin{align}
Q_2(k)F_2(k)e^{-iE^f(k)t}+Q_1(k)F_1(k)e^{iE^f(k)t}\notag\\
=\cos{E^f(k)t}+i{\hat{d}^i(k)\cdot\hat{d}^f(k)}\sin{E^f(k)t}.
\end{align}
This gives us 
\begin{align}
g_k(t)=\cos{E^f(k)t}+i{\hat{d}^i(k)\cdot\hat{d}^f(k)}\sin{E^f(k)t}.
\end{align}\\

\begin{figure*}[t]
  \centering
  {\includegraphics[width=1\textwidth]{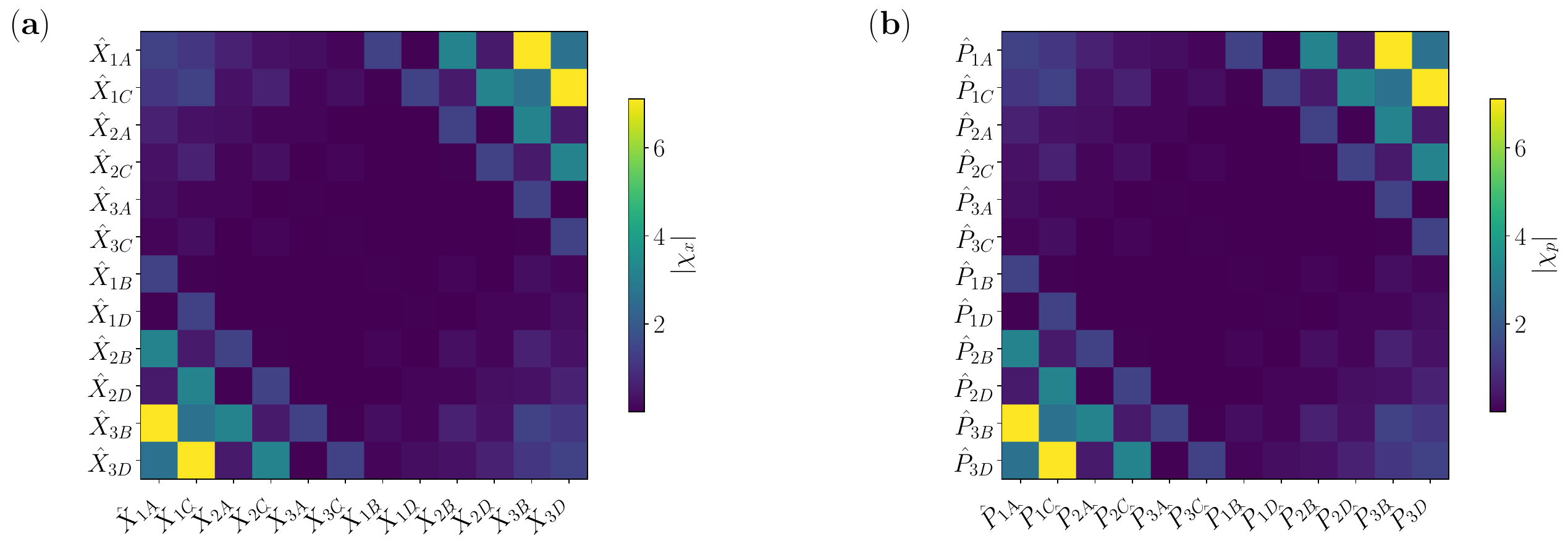}}
  \captionsetup{justification=justified,singlelinecheck=false}
  \caption{Susceptibility matrices $\chi_x$ and $\chi_p$ for finite $\gamma$, showing sublattice-dependent directional amplification. (a) $\chi_x$ and (b) $\chi_p$ show that the quadratures $\hat{X}$ and $\hat{P}$ associated with sublattices $A$ and $C$ are amplified towards left. On the contrast, the quadratures $\hat{X}$ and $\hat{P}$ associated with sublattices $B$ and $D$ are amplified towards right. The parameters used to generate these plots are: $J=1$, $\delta=0.3$, $\theta=0.4$ and $\gamma=0.03$.}
  \label{Fig_Susc_1}
\end{figure*}

\section{Amplification in the imaginary parameter regime}\label{App.susc}
In this appendix, we give the dynamical matrices $\textbf{h}_x$ and $\textbf{h}_p$ corresponding to $\hat{X}$ and $\hat{P}$ quadratures for OBC in the imaginary parameter regime and discuss the effect of dissipation $\gamma$. The Heisenberg equations for the $\hat{X}$ and $\hat{P}$ quadratures given in Eqs. \ref{Heis_X} and \ref{Heis_P} can be written in the matrix form as in Eq. \ref{xp_mat}.
\begin{align}\label{sus1}
\begin{pmatrix}
\dot{\hat{\textbf{X}}} \\
\dot{\hat{\textbf{P}}} 
\end{pmatrix}&=\begin{pmatrix}
\textbf{h}_x-\frac{\gamma}{2}\textbf{1} & \textbf{0} \\
\textbf{0} & \textbf{h}_p-\frac{\gamma}{2}\textbf{1}
\end{pmatrix}\begin{pmatrix}
\hat{\textbf{X}} \\
\hat{\textbf{P}}
\end{pmatrix}-\sqrt{\gamma}\begin{pmatrix}
\hat{\textbf{X}}^{IN} \\
\hat{\textbf{P}}^{IN}
\end{pmatrix}\notag\\
&=\textbf{M}\begin{pmatrix}
\hat{\textbf{X}} \\
\hat{\textbf{P}}
\end{pmatrix}-\sqrt{\gamma}\begin{pmatrix}
\hat{\textbf{X}}^{IN} \\
\hat{\textbf{P}}^{IN}
\end{pmatrix},
\end{align}
where $\hat{\textbf{X}}= (\hat{X}_{1A}, \hat{X}_{1B}, \hat{X}_{1C}, \hat{X}_{1D},...,\hat{X}_{NA}, \hat{X}_{NB}, \hat{X}_{NC}, \hat{X}_{ND})^T$ and similarly for $\hat{P}$. The vectors $\hat{\textbf{X}}^{IN}$ and $\hat{\textbf{P}}^{IN}$ are the corresponding input modes associated with $\hat{X}$ and $\hat{P}$ quadratures.
We specify the matrices $\textbf{h}_x$ and $\textbf{h}_p$ for system size of $N=2$.
\[
\textbf{h}_x=
\begin{bmatrix}
0 & v & 0 & 0 & 0 & 0 & 0 & 0\\
-v & 0 & 0 & 0 & -\frac{w_r+w_l}{2} & 0 & \frac{w_l-w_r}{2} & 0\\
0 & 0 & 0 & -v & 0 & 0 & 0 & 0\\
0 & 0 & v & 0 & \frac{w_l-w_r}{2} & 0 & \frac{w_r+w_l}{2} & 0\\
0 & \frac{w_r+w_l}{2} & 0 & \frac{w_l-w_r}{2} & 0 & v & 0 & 0\\
0 & 0 & 0 & 0 & -v & 0 & 0 & 0\\
0 & \frac{w_l-w_r}{2} & 0 & -\frac{w_r+w_l}{2} & 0 & 0 & 0 & -v\\
0 & 0 & 0 & 0 & 0 & 0 & v & 0
\end{bmatrix},
\]

\[
\textbf{h}_p=
\begin{bmatrix}
0 & v & 0 & 0 & 0 & 0 & 0 & 0\\
-v & 0 & 0 & 0 & -\frac{w_r+w_l}{2} & 0 & -\frac{w_l-w_r}{2} & 0\\
0 & 0 & 0 & -v & 0 & 0 & 0 & 0\\
0 & 0 & v & 0 & -\frac{w_l-w_r}{2} & 0 & \frac{w_r+w_l}{2} & 0\\
0 & \frac{w_r+w_l}{2} & 0 & -\frac{w_l-w_r}{2} & 0 & v & 0 & 0\\
0 & 0 & 0 & 0 & -v & 0 & 0 & 0\\
0 & -\frac{w_l-w_r}{2} & 0 & -\frac{w_r+w_l}{2} & 0 & 0 & 0 & -v\\
0 & 0 & 0 & 0 & 0 & 0 & v & 0
\end{bmatrix}.
\]
We take the Fourier transform of the quadrature operators as
\begin{align}
\hat{X}_{j\alpha}(t)=\frac{1}{2\pi}\int d\omega e^{-i\omega t}\hat{X}_{j\alpha}(\omega),
\end{align}
and similarly for $P$ quadratures. After the Fourier transform, Eq.~\ref{sus1} gives the susceptibility matrix as
$\chi(\omega)=(\textbf{M}+i\omega \textbf{1})^{-1}$. We are here interested in the static susceptibility at zero frequency, $\chi\equiv\chi(0)=\textbf{M}^{-1}$. 
In the $\gamma \rightarrow 0$ limit, sublattices $A$ and $C$ both couple only to sublattices $B$ and $D$, and vice versa. Consequently, the susceptibility matrices $\chi_x$ and $\chi_p$ can be reorganized into block submatrices $\chi^{AC}_{x,p}$ and $\chi^{BD}_{x,p}$ (see Fig. \ref{Susc}). However, for finite $\gamma$, such a block decomposition is no longer possible, and the full susceptibility matrices must be analyzed. The corresponding plots for finite $\gamma$ are shown in Fig. \ref{Fig_Susc_1}. We still observe amplification, but now the susceptibility matrices develop small nonzero elements between previously uncoupled sublattices, such as between $A$ and $A$, or $A$ and $C$, or others, which were absent in the $\gamma\rightarrow 0$ limit. We also note that a large $\gamma$ hinders the amplification since it introduces dissipation.\\

\section{Loschmidt echo and dynamical topological order parameter in the imaginary parameter regime}\label{LE_for_imaginary}
\begin{figure}[h]
  \centering
 {\includegraphics[width=0.49\textwidth]{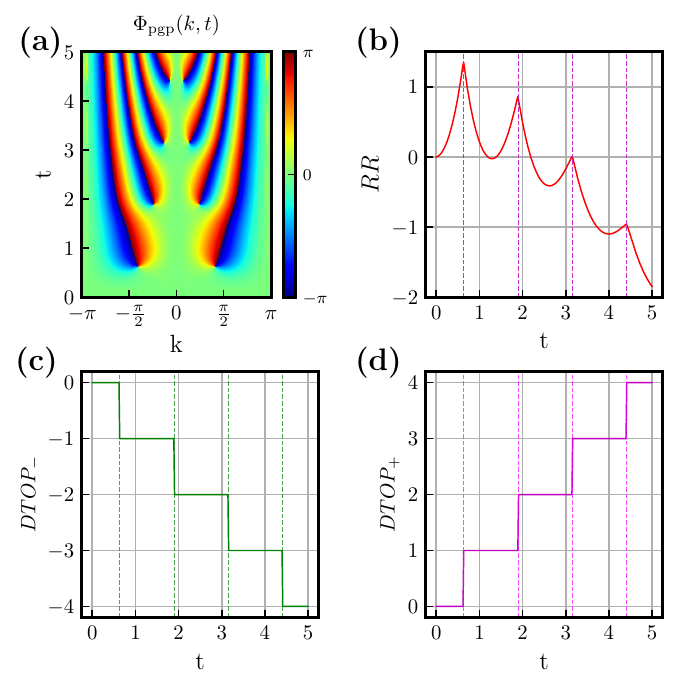}}
  \caption{Dynamical response for the imaginary parameter regime. The Pancharatnam geometric phase, $\phi_{\rm pgp}(k,t)$ in (a), the return rate ($RR(t)$) in (b), the $DTOP_-$ and $DTOP_+$ in (c) and (d), respectively, for the quench from an initial Hamiltonian with parameters $J^i=1$, $\delta^i=-1$, $\theta^i=0$ to a final Hamiltonian with parameters $J^f=1$, $\delta^f=1$, $\theta^f=0.4$.}
  \label{PGP_DTOP_imaginary_t_nt}
\end{figure}

In Sec. \ref{QBHresp}, we discussed the real parameter regime, where we showed that a chiral response appears in the PGP and the DTOP when the quench is performed across the trivial and Möbius phases. In the imaginary parameter regime, however, the Möbius phase is absent, and the response in the PGP and the DTOP are therefore expected to be symmetric in the BZ. In this Appendix, we explicitly demonstrate this symmetry. The analytical expressions of the Loschmidt amplitude and the PGP are analogous to that in the real parameter regime. The Loschmidt amplitude is given by
\begin{align}
g_k(t)=\cos{\tilde{E}^f(k)t}+i{\hat{\tilde{d}}^i(k)\cdot\hat{\tilde{d}}^f(k)}\sin{\tilde{E}^f(k)t},
\end{align}
where the superscripts $i$ and $f$ denote the quantities associated with the initial and final Hamiltonians. From this, the PGP can be obtained as 
\begin{align}
\phi_{\rm pgp}(k,t)=\phi_k(t)-\phi_{dyn}(k,t),
\end{align}
where $\phi_{dyn}(k,t)$ is the dynamical phase given by 
\begin{align}
\phi_{dyn}(k,t)=\tilde{E}^f(k)\hat{\tilde{d}}^i(k)\cdot\hat{\tilde{d}}^f(k)t.
\end{align}
The DTOP over the positive and negative halves of the BZ, denoted by $DTOP_+$ and $DTOP_-$, respectively is given by: 
\begin{align}
DTOP_+(t)&=\frac{1}{2\pi}\int_0^\pi \partial_k \phi_{\rm pgp}(k,t)dk,\notag\\
DTOP_-(t)&=\frac{1}{2\pi}\int_{-\pi}^0 \partial_k \phi_{\rm pgp}(k,t)dk.
\end{align}
Since the Möbius phase does not occur in the imaginary parameter regime, we compute the PGP, the RR and the DTOP for a quench from the trivial to the non-trivial phase. The results are shown in Fig.~\ref{PGP_DTOP_imaginary_t_nt}. 
As expected, the PGP exhibits symmetry across both halves of the BZ. The DTOP changes occur simultaneously in $DTOP_+$ and $DTOP_-$. The RR shows non-analyticities at critical times corresponding to these jumps, confirming that the dynamical response is non-chiral in the imaginary parameter regime.

\section{Amplification in the real parameter regime}\label{amp_for_real}
In the main text, we discuss the chiral topological amplification in Sec.~\ref{amplification} in imaginary parameter regime. In this appendix, we discuss what happens in the real parameter regime. We first write the Hamiltonian in Eq.~\ref{HQB_realspaceham} in the $\hat{X}$ and $\hat{P}$ quadrature basis. 
\begin{align}
&\hat{H}_{QB} = \sum_{j=1}^{N} \biggl[
 v \big( \hat{X}_{jA} \hat{X}_{jB} + \hat{P}_{jA} \hat{P}_{jB} \big)
 - v \big( \hat{X}_{jC} \hat{X}_{jD} + \hat{P}_{jC} \hat{P}_{jD} \big) \notag\\
&\quad +\frac{w_r+w_l}{2} \Big(
  \hat{X}_{j+1A} \hat{X}_{jB} + \hat{P}_{j+1A} \hat{P}_{jB}
  - \hat{X}_{j+1C} \hat{X}_{jD} - \hat{P}_{j+1C} \hat{P}_{jD} \Big) \notag\\
&\quad +\frac{w_l-w_r}{2} \Big(
  \hat{X}_{jB} \hat{X}_{j+1C} - \hat{X}_{j+1A} \hat{X}_{jD}
  - \hat{P}_{jB} \hat{P}_{j+1C} + \hat{P}_{j+1A} \hat{P}_{jD}
\Big) \biggr].
\end{align}
From this Hamiltonian, we derive the equations of motion for the $\hat{X}$ and $\hat{P}$ quadratures using the Heisenberg equations. They are given as, for the $\hat{X}$ quadratures:
\begin{align}
\dot{\hat{X}}_{jA}&=v\hat{P}_{jB}+\frac{w_r+w_l}{2}\hat{P}_{j-1B}+\frac{w_l-w_r}{2}\hat{P}_{j-1D}-\frac{\gamma}{2}\hat{X}_{jA}-\sqrt{\gamma}\hat{X}_{jA}^{IN},\notag\\
\dot{\hat{X}}_{jB}&=v\hat{P}_{jA}+\frac{w_r+w_l}{2}\hat{P}_{j+1A}-\frac{w_l-w_r}{2}\hat{P}_{j+1C}-\frac{\gamma}{2}\hat{X}_{jB}-\sqrt{\gamma}\hat{X}_{jB}^{IN},\notag\\
\dot{\hat{X}}_{jC}&=-v\hat{P}_{jD}-\frac{w_r+w_l}{2}\hat{P}_{j-1D}-\frac{w_l-w_r}{2}\hat{P}_{j-1B}-\frac{\gamma}{2}\hat{X}_{jC}-\sqrt{\gamma}\hat{X}_{jC}^{IN},\notag\\
\dot{\hat{X}}_{jD}&=-v\hat{P}_{jC}-\frac{w_r+w_l}{2}\hat{P}_{j+1C}+\frac{w_l-w_r}{2}\hat{P}_{j+1A}-\frac{\gamma}{2}\hat{X}_{jD}-\sqrt{\gamma}\hat{X}_{jD}^{IN},\notag\\
\end{align}
and for the $\hat{P}$ quadratures:
\begin{align}
\dot{\hat{P}}_{jA}&=-v\hat{X}_{jB}-\frac{w_r+w_l}{2}\hat{X}_{j-1B}+\frac{w_l-w_r}{2}\hat{X}_{j-1D}-\frac{\gamma}{2}\hat{P}_{jA}-\sqrt{\gamma}\hat{P}_{jA}^{IN},\notag\\
\dot{\hat{P}}_{jB}&=-v\hat{X}_{jA}-\frac{w_r+w_l}{2}\hat{X}_{j+1A}-\frac{w_l-w_r}{2}\hat{X}_{j+1C}-\frac{\gamma}{2}\hat{P}_{jB}-\sqrt{\gamma}\hat{P}_{jB}^{IN},\notag\\
\dot{\hat{P}}_{jC}&=v\hat{X}_{jD}+\frac{w_r+w_l}{2}\hat{X}_{j-1D}-\frac{w_l-w_r}{2}\hat{X}_{j-1B}-\frac{\gamma}{2}\hat{P}_{jC}-\sqrt{\gamma}\hat{P}_{jC}^{IN},\notag\\
\dot{\hat{P}}_{jD}&=v\hat{X}_{jC}+\frac{w_r+w_l}{2}\hat{X}_{j+1C}+\frac{w_l-w_r}{2}\hat{X}_{j+1A}-\frac{\gamma}{2}\hat{P}_{jD}-\sqrt{\gamma}\hat{P}_{jD}^{IN}.\notag\\
\end{align}
We observe that, irrespective of the strength of dissipation, the $\hat{X}$ and $\hat{P}$ quadratures do not evolve independently in contrast to the imaginary parameter regime (in Eqs.~\ref{Heis_X} and \ref{Heis_P}). This coupling implies that the $\hat{X}$ and $\hat{P}$ quadratures do not form independent dynamical sectors in the real parameter regime. The time evolution of each quadrature explicitly depends on the other, and therefore their dynamics cannot be disentangled into separate amplification channels. In contrast to the imaginary parameter regime, where the directional amplification can be discussed for each quadrature, such a distinction is not meaningful here. Therefore, the coupling between $\hat{X}$ and $\hat{P}$ quadratures in the equations of motion in the real parameter regime precludes any well-defined notion of $\hat{X}$ or $\hat{P}$ quadrature amplification.
\bibliography{references}

\end{document}